%% file: sfrl_rev.tex
\documentclass[english,onecolumn]{IEEEtran}
\usepackage[T1]{fontenc}
\usepackage{color}
\usepackage{amsmath}
\usepackage{amsthm}
\usepackage{amssymb}
\usepackage{graphicx}
\usepackage{esint}

\makeatletter
\theoremstyle{plain}
\newtheorem{thm}{\protect\theoremname}
\theoremstyle{definition}
\newtheorem{defn}{\protect\definitionname}
\theoremstyle{definition}
\newtheorem{rem}{\protect\remarkname}
\theoremstyle{plain}
\newtheorem{prop}{\protect\propositionname}
\theoremstyle{definition}
\newtheorem{example}{\protect\examplename}

\include{defns}

\renewcommand{\E}{\normalfont\textsf{E}}
\renewcommand{\P}{\normalfont\textsf{P}}
\DeclareMathOperator\indep{{\perp\!\!\!\perp}}

\usepackage{mathrsfs}

\makeatother

\usepackage{babel}
\providecommand{\definitionname}{Definition}
\providecommand{\propositionname}{Proposition}
\providecommand{\remarkname}{Remark}
\providecommand{\theoremname}{Theorem}
\providecommand{\examplename}{Example}

\begin{document}

\title{Strong Functional Representation Lemma and Applications to Coding Theorems}

\author{Cheuk Ting Li and Abbas El Gamal\\
 Department of Electrical Engineering, 
 Stanford University\\
 Email: ctli@stanford.edu, abbas@ee.stanford.edu \\
\thanks{This paper was presented in part at the IEEE International
Symposium on Information Theory, Aachen, Germany, June 2017.%
}
}
%\author{Cheuk Ting Li, {\it Student Member, IEEE}, and Abbas El Gamal, {\it Fellow, IEEE}\\
% \thanks{This paper was presented in part at the IEEE International
%Symposium on Information Theory, Aachen, Germany, June 2017.%
%
%The authors are with the Department of Electrical Engineering, Stanford University, Stanford, CA 94305, USA (e-mail: \mbox{ctli@stanford.edu}, \mbox{abbas@ee.stanford.edu}).
%}
%}
   
\maketitle
\begin{abstract}
This paper shows that for any random variables $X$ and $Y$, it is possible to represent $Y$ as a function of $(X,Z)$ such that $Z$ is independent of $X$ and $I(X;Z|Y)\le\log(I(X;Y)+1)+4$ bits.
%We use this strong functional representation lemma (SFRL) to establish a tighter bound on the rate needed for one-shot exact channel simulation than was previously established by Harsha {\it et al.},
\textcolor{black}{We use this strong functional representation lemma (SFRL) to establish a bound on the rate needed for one-shot exact channel simulation for general (discrete or continuous) random variables, strengthening the results by Harsha {\it et al.} and Braverman and Garg,} and to establish new and simple achievability results for one-shot variable-length lossy source coding, multiple description coding and Gray-Wyner system. We also show that the SFRL can be used to reduce the channel with state noncausally known at the encoder to a point-to-point channel, which provides a simple achievability proof of the Gelfand-Pinsker theorem.
%Finally we present an example in which the SFRL inequality is tight to within 5 bits.
\end{abstract}

\section{Introduction}

The functional representation lemma~\cite[p. 626]{elgamal2011network} states that for any random variables $X$ and $Y$, there exists a random variable $Z$ independent of $X$ such that $Y$ can be represented as a function of $X$ and $Z$. This result has been used to establish several results in network information theory beginning with the early work of Hajek and Pursley on the broadcast channel~\cite{hajek1979} and Willems and van der Meulen on the multiple access channel with cribbing encoders~\cite{willems1985}.

%%%AEG 
\textcolor{black}{
The random variable $Z$ in the functional representation lemma can be intuitively viewed as the part of $Y$ which is not contained in $X$. However, $Z$ is not necessarily unique. For example, let $B_1,B_2,B_3,B_4$ be i.i.d. $\Bern(1/2)$ random variables and define $X=(B_1,B_2,B_3)$ and $Y=(B_2,B_3,B_4)$. Then both $Z_1=B_4$ and $Z_2= B_1\oplus B_4$ satisfy the functional representation lemma. However, $H(Y|Z_1)=2$ while $H(Y|Z_2)=3$, that is, $Z_1$ provides more information about $Y$ than $Z_2$. In general, $H(Y|Z) = I(X;Y|Z) + H(Y|X,Z) = I(X;Y,Z) \ge I(X;Y)$. For our example $H(Y|Z_1)=I(X;Y)=2$, that is, $Z_1$ is the most informative $Z$ about $Y$. What is the most informative $Z$ about $Y$ in general? Does it always achieve the lower bound $H(Y|Z) \ge I(X;Y)$?
}

\textcolor{black}{
%In this paper, we establish a general upper bound on $H(Y|Z)$ and provide a construction of $Z$ that satisfies this upper bound for general $(X,Y)$.
In this paper, we show that for general $(X,Y)$, their exists a $Z$ such that $H(Y|Z)$ is close to $I(X;Y)$.
Specifically, we strengthen the functional representation lemma to show that for any $X$ and $Y$, there exists a $Z$ independent of $X$ such that $Y$ is a function of $X$ and $Z$, \emph{and}
\[
I(X;Z|Y) \le \log(I(X;Y)+1)+4.
\]
Alternatively this can be expressed as
\begin{align}
H(Y|Z) &\le I(X;Y) + \log(I(X;Y)+1)+4. \label{eqn:sfrl_hyz}
\end{align}
}
%%%AEG

We use the above {\em strong functional representation lemma} (SFRL) together with an optimal prefix code such as a Huffman code to establish one-shot, \emph{variable-length} \textcolor{black}{achievability} results for channel simulation~\cite{harsha2010communication}, Shannon's lossy source coding~\cite{Shannon1959}, multiple description coding~\cite{El-Gamal--Cover1982,Zhang--Berger1987} and lossy Gray--Wyner system~\cite{Gray--Wyner1974}. \textcolor{black}{These one-shot achievability results can be stated in terms of mutual information, without the need of information density or other quantities.} We then show how the SFRL can be used to reduce the channel with state known at the encoder to a point-to-point channel, providing a simple proof to the Gelfand-Pinsker theorem~\cite{Gelfand--Pinsker1980a}. The asymptotic block coding counterparts of these one-shot results can be readily obtained by converting the variable-length code into a block code and incurring an error probability that vanishes as the block length approaches infinity.

\textcolor{black}{
A weaker form of the SFRL for discrete random variables follows from the result by Harsha {\it et al.}~\cite{harsha2010communication} on the one-shot exact channel simulation with unlimited common randomness. Their result  implies that
$I(X;Z|Y)\le{(1+\e)}\log(I(X;Y)+1)+c_{\e}$
is achievable, where $\e>0$ and $c_{\e}$ is a function of $\e$.
This result was later strengthened by Braverman and Garg~\cite{braverman2014public} to $I(X;Z|Y)\le\log(I(X;Y)+1)+c$ (note that replacing the universal code in~\cite{harsha2010communication} by a code for a suitable power law distribution can also yield the same improvement). It is also shown in~\cite{braverman2014public} that there exist examples for which the $\log$ term is necessary.
% However, the results in~\cite{harsha2010communication,braverman2014public} only works for discrete $Y$, and the constants are unspecified, which is undesirable in one-shot results.
SFRL strengthens these results in two ways; first it generalizes the bound to random variables with arbitrary distributions (whereas the results in~\cite{harsha2010communication,braverman2014public} only applies to discrete distributions), and second it provides a bound with a small additive constant of 4 (whereas the constants in~\cite{harsha2010communication,braverman2014public} are unspecified). Our stronger result is established using a new construction of $Z$ and $g$ that we refer to as the
\emph{Poisson functional representation}, instead of the rejection sampling approach in~\cite{harsha2010communication,braverman2014public}.
Perhaps more importantly, we are the first to show that the result in~\cite{harsha2010communication} can be considered as a strengthened functional representation lemma, which led us to explore applications in source and channel coding.
}

%A weaker form of the SFRL for discrete random variables can be obtained using the result by Harsha {\it et al.}~\cite{harsha2010communication} on the one-shot exact channel simulation with unlimited common randomness. Assuming the input $X$ has a given pmf, then~\cite{harsha2010communication} implies that
%$I(X;Z|Y)\le{(1+\e)}\log(I(X;Y)+1)+c_{\e}$
%is achievable, where $\e>0$ and $c_{\e}$ is a function of $\e$. SFRL strengthens this result in two way; first it provides a tighter bound, and second it generalizes the bound to random variables with arbitrary distributions. This is obtained via a new construction of $Z$ that we refer to as the
%\emph{Poisson functional representation} instead of the rejection sampling approach in~\cite{harsha2010communication}.
%Perhaps more importantly, we are the first to show that the result in~\cite{harsha2010communication} can be considered as a strengthened functional representation lemma, which led us to explore applications in source and channel coding.

One-shot achievability results using fixed length (random) coding have been recently established for lossy source coding and several settings in network information theory. In~\cite{liu2015resolvability}, Liu, Cuff and Verd\'u established a one-shot achievability result for lossy source coding using channel resolvability. One-shot quantum lossy source coding settings were investigated by Datta {\it et al.}~\cite{datta2013quantum}.
%that were previously studied only in the asymptotic regime.
In~\cite{verdu2012nonasymp}, Verd\'u  introduced non-asymptotic packing and covering lemmas and used them
to establish one-shot achievability results for several settings including Gelfand-Pinsker.
In~\cite{liu2015oneshot}, Liu, Cuff
and Verd\'u proved a one-shot mutual covering lemma and used it to establish 
a one-shot achievability result for the broadcast channel.
%with private and common messages that subsumes Marton's inner bound.
In~\cite{watanabe2013sideinfo}, Watanabe, Kuzuoka and Tan established several one-shot achievability results for coding with side-information (including Gelfand-Pinsker).
In~\cite{yassaee2013oneshot},
Yassaee, Aref and Gohari established several one-shot achievability
results, including Gelfand-Pinsker and multiple description coding. Most of these results are stated in terms of information density and various other quantities. In contrast, our one-shot achievability results using variable-length codes are all stated in terms of only mutual information. Moreover, given the SFRL, our proofs are generally simpler.

Variable-length (one-shot, finite blocklength or asymptotic) lossy source coding settings have been studied, e.g., see~\cite{pinkston1967encoding,pursley1976varrate,mackenthun1978varrate,kosut2013variable,kostina2015variable}.
Some of these works concern the universal setting in which the distribution of the source is unknown, hence the use of variable-length codes is justified. In contrast, the reason we consider variable-length codes in this paper is that it allows us to give one-shot results that subsume their asymptotic fixed-length counterparts.

\textcolor{black}{In the following section, we state the SFRL, introduce the Poisson functional representation construction and provide a sketch of the proof of the lemma. The complete proof is given in Appendix~\ref{subsec:strongfrl}.} In Sections~\ref{secchansim} and~\ref{sec:onesrccode} we use SFRL to establish one-shot achievability results for channel simulation and three source coding settings, respectively. In Section~\ref{sec:state}, we use SFRL together with Shannon's channel coding theorem to provide a simple achievability proof of the Gelfand--Pinsker theorem.
%In Section~\ref{sec:cont} we extend the proof of the SFRL to general random variables.
%Finally in Section~\ref{sec:tightness} we demonstrate the tightness of the SFRL inequality and discuss several other properties of this inequality.
\textcolor{black}{Finally in Section~\ref{sec:tightness} we prove a lower bound on $I(X;Z|Y)$ in SFRL (whereas SFRL is an upper bound) and discuss several other properties.}

\subsection*{Notation}

Throughout this paper, we assume that $\log$ is base 2 and the entropy
$H$ is in bits.
% The binary entropy function is \textcolor{black}{$H(p)=-p\log p-(1-p)\log(1-p)$}.
We use the notation: $X_{a}^{b}=(X_{a},\ldots,X_{b})$,
$X^{n}=X_{1}^{n}$, $[a:b]=[a,b]\cap\mathbb{Z}$ and
$[a]=[1:a]$.

For discrete $X$, we write the probability mass function
as $p_{X}$. For continuous $X$, we write the probability density
function as $f_{X}$. For general random variable $X$, we write the
probability measure (push-forward measure by $X$) as $\P_{X}$. 
%For jointly distributed $\P_{X,Y}$, the \emph{lautum information}~\cite{palomar2006}
%is defined as
%\[
%L(X;Y)=D_{\mathrm{KL}}(\P_{X}\P_{Y}\Vert\P_{XY}).
%\]

\section{Strong Functional Representation Lemma\label{sec:strongfrl}}

The main result in this paper is given in the following.
\begin{thm}
[Strong functional representation lemma]\label{thm:strongfrl} For any pair of random variables
$(X,Y) \sim \P_{XY}$ (over a Polish space with Borel probability
measure) with $I(X;Y)<\infty$, there exists a random variable $Z$ independent of $X$ such that $Y$ can be expressed as a function $g(X,Z)$ of $X$ and $Z$, and
\[
I(X;Z|Y)\le\log(I(X;Y)+1)+4.
\]
Moreover, if $X$ and $Y$ are discrete with cardinalities $|\Xc|$ and $|\Yc|$, respectively, then
 $\left|\Zc \right|\le |\Xc|(|\Yc|-1)+2$.
\end{thm}
Note that SFRL can be applied conditionally; given $\P_{XY|U}$, we can represent $Y$ as a function $g(X,Z,U)$ such that
$Z$ is independent of $(X,U)$ and
\begin{align}
I(X;Z|Y,U)\le\log\left(I(X;Y|U)+1\right)+4. \label{eqn:sfrl_cond}
\end{align}
\textcolor{black}{We can have $Z\indep (X,U)$, not only $Z\indep X\, |\, U$ which follows from directly applying SFRL for each value of $U$. The reason is that by the functional representation lemma, we can represent $Z$ as a function of $U$ and $\tilde{Z}$ such that $\tilde{Z} \indep U$ (which, together with $\tilde{Z}\indep X\, |\, U$, gives $\tilde{Z} \indep (X,U)$), and use $\tilde{Z}$ instead of $Z$.}

Note that SFRL applies to general distributions $\P_{XY}$.
% Although $H(Y)$ may be infinite, the cardinality of $Y$ conditioned on $Z$
% can still be countable and $H(Y|Z)$ can be finite.
\textcolor{black}{Although $H(Y)$ may be infinite, as long as $I(X;Y)$ is finite, the cardinality of $Y$ conditioned on $Z$
is countable and $H(Y|Z)$ is finite by SFRL.}
Since $Z\indep X$ and $H(Y|X,Z)=0$ imply that $I(X;Z|Y)=H(Y|Z)-I(X;Y)$, the SFRL implies the existence of a $Z\indep X$ such that $H(Y|Z)$ is close to $I(X;Y)$.
\medskip

{\color{black}
To prove the SFRL, we use the following random variable $Z$ and function $g$ construction.
\begin{defn}
[Poisson functional representation] Fix any joint distribution $\P_{XY}$.
Let $0\le T_{1}\le T_{2}\le \cdots$ be a Poisson point process with
rate 1 (i.e., the increments $T_{i}-T_{i-1}$ are i.i.d. $\Exp(1)$ for $i=1,2,\ldots$ with $T_{0}=0$), and $\tilde{Y}_{1},\tilde{Y}_{2},\ldots$ be i.i.d.  with $\Yt_1 \sim \P_{Y}$. Take $Z=\{(\tilde{Y}_{i}, T_{i})\}_{i=1,2,\ldots}$, i.e., a marked
Poisson point process. Then we can let $Y=g_{X\to Y}(X,Z)$, where
\[
g_{X\to Y}(x,\,\{(\tilde{y}_{i}, t_{i})\})=\tilde{y}_{k_{X\to Y}(x,\,\{(\tilde{y}_{i}, t_{i})\})},
\]
and
\[
k_{X\to Y}(x,\,\{(\tilde{y}_{i}, t_{i})\})=\underset{i}{\arg\min}\,\,t_{i}\cdot\frac{d\,\P_{Y}}{d\,\P_{Y|X}(\cdot|x)}(\tilde{y}_{i}).
\]
\end{defn}

To illustrate this Poisson functional representation, consider the following.

\begin{example}
Let $Y \sim \U[0,1]$ and $Y|\{X=x\} \sim f_{Y|X}(y|x)$. Then $g_{X \to Y}(x, z) = \tilde{y}_k$ where $k = \arg \min_i t_i / f_{Y|X}(y|x)$. Figure~\ref{fig:poisson} shows an example of $z=\{(\tilde{y}_{i}, t_{i})\}$. The index $k$ is selected by scaling up the graph of $f_{Y|X}(y|x)$ until it hits the first point, then we output $\tilde{y}_k$ of that point ($\tilde{y}_3$ in the figure). It is straightforward to check that this procedure gives the correct conditional distribution $Y|\{X=x\} \sim f_{Y|X}(y|x)$. Roughly speaking, if $I(X;Y)$ is small, then $Y|\{X=x\}$ will be close to the uniform distribution for most $x$'s, and the $\tilde{y}_k$'s with smaller indices $k$'s will be more likely to be output, and therefore $H(Y|Z)$ will be smaller. (If $I(X;Y)=0$, then $Y|\{X=x\} \sim \U[0,1]$ and $\tilde{y}_1$ is output for almost all $x$, and hence $H(Y|Z)=0$.)
\end{example}

\begin{figure}[h]
\begin{center}
\includegraphics[scale=0.75]{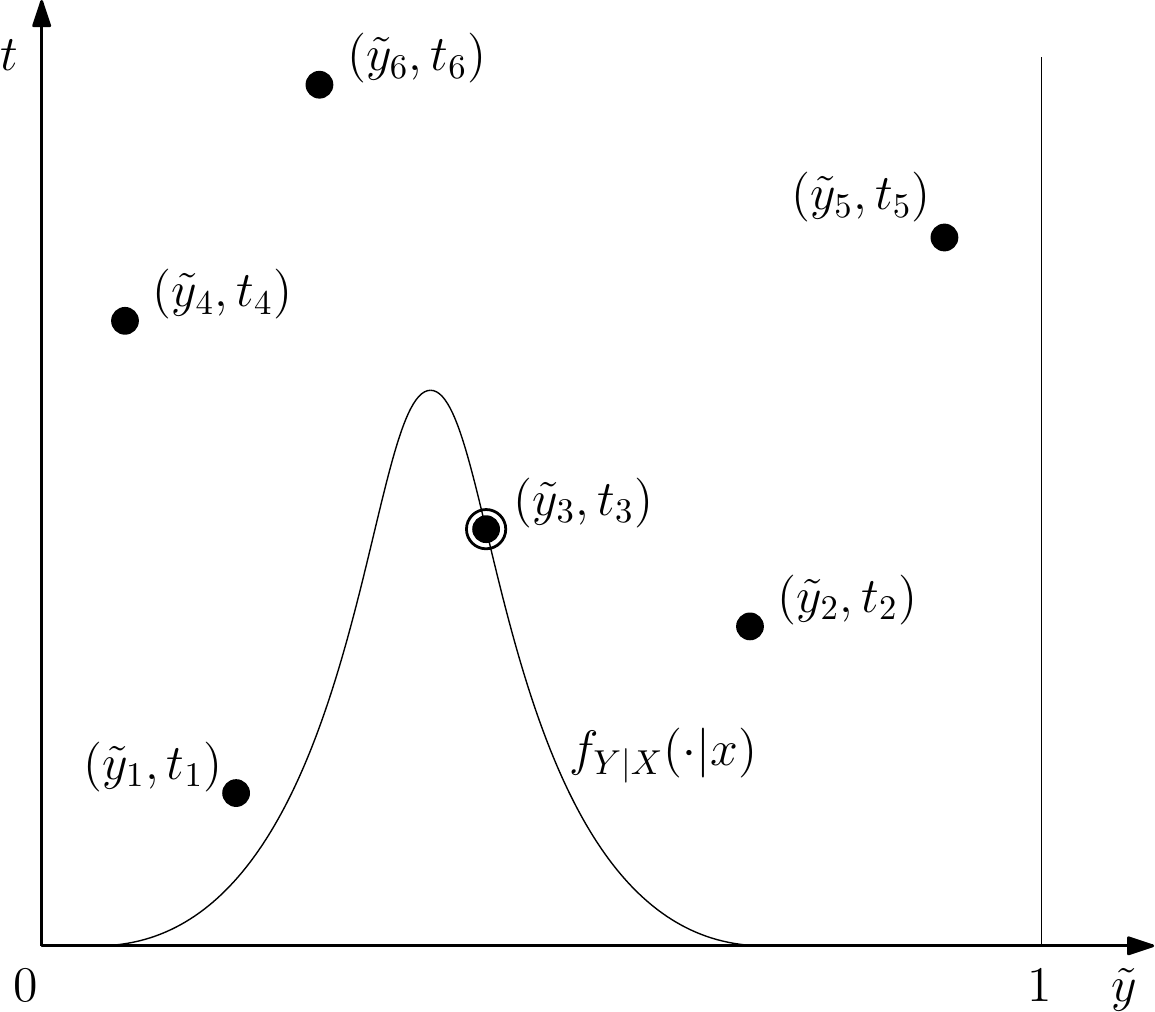}
\caption{\label{fig:poisson}Illustration of the Poisson functional representation construction for Example 1.}
\end{center}
\end{figure}

\begin{rem}
If $Y$ is discrete, then we can simplify the construction of $Z$ to a vector of exponential random variables by letting $Z_{y}=p_{Y}(y)\cdot\min_{i:\,\tilde{Y}_{i}=y}T_{i}$, which leads to the following construction. Assume $Y\in\{1,\ldots,l\}$. We can let $Y=g_{X\to Y}(X,Z^{l})$,
where $Z^{l}$ is a sequence of i.i.d. $\Exp(1)$ random variables independent of $X$, and
\[
g_{X\to Y}(x,z^{l})=\underset{y\in\mathcal{Y}}{\arg\min}\frac{z_{y}}{p_{Y|X}(y|x)}.
\]
\end{rem}}
\medskip
%Since the $\arg \min$ of independent exponential random variables with different rates has a pmf proportional to the rates, we have $g_{X\to Y}(x,Z^{\left|\mathcal{Y}\right|}) \sim p_{Y|X}(\cdot|x)$.

%Note that if $|\mathcal{Y}|$ is finite, we can generate $Z^{\left|\mathcal{Y}\right|}$
%uniformly over the probability simplex on $\mathcal{Y}$. This is equivalent
%to the original scheme after normalization such that $\sum_{y}Z_{y}=1$.
%We now proceed to prove Theorem~\ref{thm:strongfrl} for discrete $Y$ by showing that
%the exponential functional representation satisfies the constraints. The proof of the general case is given in ?????????????????.
{\color{black}
We now proceed to give a sketch of the proof of Theorem~\ref{thm:strongfrl} by showing that
the Poisson functional representation satisfies the constraints. The complete proof is given in Appendix~\ref{subsec:strongfrl}.
\begin{IEEEproof}[Sketch of the proof of Theorem~\ref{thm:strongfrl}]
Consider the Poisson functional representation. Let $Y=\tilde{Y}_K$,
\[
K=k_{X\to Y}(X,\,\{(\tilde{Y}_{i}, T_{i})\})=\underset{i}{\arg\min}\,\,T_{i}\cdot\frac{d\,\P_{Y}}{d\,\P_{Y|X}(\cdot|X)}(\tilde{Y}_{i}).
\]
Since $Y$ is a function of $Z$
and $K$, we have $H(Y|Z)\le H(K)$. We now proceed to bound $H(K)$.

Condition on $X=x$. Since $T_1 \le T_2 \le \cdots$, $K$ is small when $d\,\P_{Y}(y)/d\,\P_{Y|X}(y|x)$ for different $y$'s are close to 1, i.e., $\P_{Y}$ is close to $\P_{Y|X}(\cdot|x)$ (if $\P_{Y}=\P_{Y|X}(\cdot|x)$ for all $y$, then $d\,\P_{Y}(y)/d\,\P_{Y|X}(y|x)=1$, and $K=1$). In fact we can prove that
\[
\E\left[\log K  |  X=x\right] \le D (\P_{Y|X}(\cdot|x) \Vert\, \P_Y)+e^{-1}\log e+1.
\]
The proof is given in Appendix~\ref{subsec:strongfrl}. Therefore $\E\left[\log K\right]\le I(X;Y)+e^{-1}\log e+1$. By the maximum entropy distribution subject to a given $\E\left[\log K\right]$,
we have
\[
H(K)\le\E\left[\log K\right]+\log\left(\E\left[\log K\right]+1\right)+1.
\]
The proof of this bound is given in Appendix~\ref{subsec:boundent}
for the sake of completeness.
Hence
\[
\begin{aligned}H(K) & \le I(X;Y)+e^{-1}\log e+2+\log\left(I(X;Y)+e^{-1}\log e+2\right)\\
 & \le I(X;Y)+\log\left(I(X;Y)+1\right)+e^{-1}\log e+2+\log\left(e^{-1}\log e+2\right)\\
 & <I(X;Y)+\log\left(I(X;Y)+1\right)+4.
\end{aligned}
\]
Operationally, $K$ can be encoded using the optimal prefix-free code for the Zipf distribution
$q(k)\propto k^{-\lambda}$, where
\begin{align}
\lambda=1+1/(I(X;Y)+e^{-1}\log e+1). \label{eqn:zipf}
\end{align}
It can be checked that the expected length of the codeword is upper bounded by $I(X;Y)+\log\left(I(X;Y)+1\right)+5$.

%The cardinality bound $\left|\mathcal{Z}\right|\le\left|\mathcal{Y}\right|^{\left|\mathcal{X}\right|}$
%follows directly from the fact that there are $\left|\mathcal{Y}\right|^{\left|\mathcal{X}\right|}$
%different possible functions $x\mapsto g_{X\to Y}(x,z)$ for different
%$z$, and hence the $z$'s with the same function can be merged.
\end{IEEEproof}
}

%{\color{black}
%\begin{thm}
%[General strong functional representation lemma]\label{thm:gstrongfrl} For any pair of random variables
%$(X,Y) \sim \P_{XY}$ with $I(X;Y)<\infty$, there exists a random variable $Z$ independent of $X$ such that $Y$ can be expressed as a function $g(X,Z)$ of $X$ and $Z$, and for any $A \subseteq \Xc$,
%\[
%H(Y|Z, X\in A) \le I(X;Y|X \in A) + \log(I(X;Y|X \in A)+1)+4.
%\]
%As a result, for any $W$ that is a function of $X$,
%\[
%H(Y|Z, W) \le I(X;Y|W) + \log(I(X;Y|W)+1)+4.
%\]
%\end{thm}
%}

%\begin{rem}
%\textcolor{black}{Note that if we do not fix a distribution for $X$ (but still fix a conditional distribution $p_{Y|X}$), and replace $p_Y$ with an arbitrary distribution $q_Y$, then we have the following result:}
%
%\textcolor{black}{
%For any $\P_{Y|X}$ and distribution $\textsf{Q}_Y$, there exists a random variable $Z$, conditional distribution $\textsf{Q}_{Y|Z}$ and function $g(x,z)$ such that $Y=g(x,Z) \sim p_{Y|X}(\cdot|x)$ for all $x$, and
%\[
%\E\left[L(M)\right]\le C+\log(C+1)+5
%\]
%}
%\end{rem}

%CODEWORD VS DESCRIPTION VS INDEX. WE NEED TO BE CONSISTENT

\section{One-shot Channel Simulation \label{secchansim}}

Channel simulation aims to find the minimum amount of communication over a noiseless channel needed to simulate a memoryless channel $\P_{Y|X}$. Several settings of this problem have been studied, e.g., see~\cite{bennett2002entanglement,cuff2013distributed,bennet2014reverse}. Consider the one-shot channel simulation with unlimited common randomness setup~\cite{harsha2010communication} in which Alice and Bob share unlimited common randomness $W$. Alice observes $X\sim \P_X$ and sends a prefix-free description $M$ to Bob via a noiseless channel such that Bob can generate $Y$ (from $M$ and $W$) according to a prescribed conditional distribution $\P_{Y|X}$. The problem is to find  the minimum expected description length of $M$, $\E\left[L(M)\right]$, needed.
%It is straightforward to show that  $\E\left[L(M)\right]\ge I(X;Y)$.
\textcolor{black}{Since we have the Markov chain $X-M-Y$ conditional on $W$,
\[
\E\left[L(M)\right]\ge H(M|W) \ge I(X;Y|W) = I(X;Y,W)-I(X;W) = I(X;Y,W) \ge I(X;Y).
\]
}
\textcolor{black}{In~\cite{braverman2014public}, which strengthens the result in~\cite{harsha2010communication}, it is shown that for $X$ and $Y$ discrete,
\[
\E\left[L(M)\right]\le I(X;Y)+\log(I(X;Y)+1)+c
\]
is achievable, where $c$ is an unspecified constant.}
% It is possible to remove $\e$ by using a code for Zipf distribution instead of a universal code in~\cite{harsha2010communication}, though the unspecified constant in~\cite{harsha2010communication} appears to be large.

We now show that the SFRL provides an upper bound on $\E\left[L(M)\right]$ that applies to arbitrary (not only discrete) channels. By the SFRL~\eqref{eqn:sfrl_hyz}, there exists a $Z$ independent of $X$
such that $Y=g_{X\to Y}(X,Z)$ and
\[
H(Y|Z)\le I(X;Y)+\log(I(X;Y)+1)+4.
\]
\textcolor{black}{
We use $W=Z$ as the common randomness. Upon observing $X=x$, Alice
computes $y = g_{X \to Y}(x, z)$ and encodes $y$ using a Huffman code
for the pmf $p_{Y|Z}(\cdot|z)$ into the description
$m$ (note that $Y$ can be arbitrary but by the SFRL $Y|\{Z=z\}$ is discrete). Bob then recovers $y$ from $m$ and $z$. The expected length is
\[
\E\left[L(M)\right]\le I(X;Y)+\log(I(X;Y)+1)+5.
\]
In practice, instead of using a Huffman code (which may be impractical since $p_{Y|Z}(\cdot|z)$ is not easy to compute), we can compress $k=k_{X\to Y}(x,z)$ in the Poisson functional representation into $m$ using the optimal
prefix-free code for the Zipf distribution~\eqref{eqn:zipf}.
}
%In Section~\ref{sec:tightness}, we give an
%example which shows that the $\log(I(X;Y)+1)$ term is in general necessary.

Moreover, for discrete $X,Y$, the amount of the common randomness can be bounded by  $\log |\mathcal{W} |\le \log(|\Xc|(|\Yc|-1)+2)$. In comparison, the amount of the common randomness in~\cite{harsha2010communication} can be bounded by $O(\log (|\Xc||\Yc|))$ only if the expected description length is increased by $O(\log \log (|\Xc|+|\Yc|))$.

%We can use the exponential
%functional representation to construct $W=Z^{|\mathcal{Y}|}$ (if $|\mathcal{W}|$ is unlimited). Upon observing $W$ and $X$, Alice generates
%$Y\sim\P_{Y|X}$ and finds $K$, the index of $Z_{Y}/p_{Y}(Y)$ in the set
%$\{Z_{y}/p_{Y}(y)\}_{y\in\mathcal{Y}}$ sorted in ascending order
%as in the proof of the SFRL, and encodes $K$ into $M$ using the optimal
%prefix-free code for the Zipf distribution $q(k)\propto k^{-\lambda}$,
%$k=1,2,\ldots$, where $\lambda=1+1/(I(X;Y)+e^{-1}\log e+1)$. Bob
%recovers $K$ from $M$ and obtains $Y$ using $K$ and $Z^{|\mathcal{Y}|}$.
\medskip{}

\begin{rem}
In~\cite{harsha2010communication}, the setting in which $X=x$ is
an arbitrary input (instead of $X\sim p_{X}$) is studied. It is shown
that 
\[
\E\left[L(M)\right]\le C+(1+\epsilon)\log(C+1)+c_{\e}
\]
for all $x\in\mathcal{X}$ is achievable, where $C$ is the capacity
of the channel $p_{Y|X}$ and $c_{\e}$ is a function of $\e$.

The Poisson functional representation can still be applied to this
setting. If we encode $k = k_{X \to Y}(x,z)$ into $M$
using the optimal prefix-free code for the Zipf distribution
$q(k)\propto k^{-\lambda}$, where $\lambda=1+1/(C+e^{-1}\log e+1)$,
then by the same argument in the proof of the SFRL, and Claim 3.1
in \cite{harsha2010communication}, 
\[
\E\left[L(M)\right]\le C+\log(C+1)+5
\]
is achievable.

% \textcolor{black}{
We can also prove a cardinality bound of the common randomness $Z$ in this setting.
Applying Carath\'eodory's theorem on the $(|\Xc||\Yc|)$-dimensional
vectors with entries $\E [ \log K | X=x, Z=z]$ and $p(x,y|z)$ for $x\in \{1,\ldots,|\Xc|\}$,  $y\in \{1,\ldots,|\Yc|-1\}$, we have the cardinality bound $|\Zc| \le |\Xc||\Yc| + 1$.
% }
\end{rem}
%---------------------------------------
\section{Lossy Source Coding \label{sec:onesrccode}}
We use the SFRL to establish one-shot achievability results for three lossy source coding settings.

\subsection{Lossy source coding}
Consider the following one-shot variable-length lossy source coding problem. We are given a random variable (source) $X \in \Xc$ with $X\sim \P_X$, a reproduction alphabet $\Yc$, and a distortion function $d:\Xc\times \Yc \to [0,\infty]$ (note that $X,Y$ can be arbitrary, and $d(x,y)$ can be infinite).
%We use an optimal prefix-free code to encode $X$ into an index $M \in \{0,1\}^*$. A decoder maps $M$ into $\tilde{Y} \in \Yc$.
Given $X$, the encoder selects $\tilde{Y} \in \Yc$ and encodes it using a prefix-free code into $M \in \{0,1\}^*$. The decoder recovers $\tilde{Y}$ from $M$.
Let $\bar R=\E[L(M)]$ be the expected value of the length of the description $M$ and $\E[d(X,\tilde{Y})]$ be the average distortion of representing $X$ by $\tilde{Y}$. An expected length-distortion pair $(\bar R,D)$ is said to be achievable if there exists a variable-length code with expected description length $\bar R$ such that $\E[d(X,\tilde{Y})] \le D$.
%Note that variable-length schemes for lossy source coding have been widely studied (e.g.~\cite{pinkston1967encoding,pursley1976varrate,ornstein1990universal}).

In the following we use the SFRL to establish a set of achievable $(\bar R,D)$ pairs.
\begin{thm}
\label{thm:oneshotlossy}
A pair $(\bar R,D)$ is achievable for the one-shot variable-length lossy source coding problem with source $X\sim \P_X$, reproduction alphabet $\Yc$, and distortion measure $d(x,y)$ if
\begin{align*}
\bar R &> R(D)+\log(R(D)+1)+6,
\end{align*}
where
\begin{align*}
R(D) &= \inf_{\P_{Y|X}:\,\E[d(X,Y)] \le D} I(X;Y)
\end{align*}
is the (asymptotic) rate-distortion function~\cite{Shannon1959}.
\end{thm}
\begin{IEEEproof}
Let $Y$ be the random variable that attains $\E[d(X,Y)] \le D$ and $I(X;Y)\le R(D) +\epsilon$.
By the SFRL~\eqref{eqn:sfrl_hyz}, there exists $Z$ independent of $X$ such that
$Y=g_{X\to Y}(X,Z)$ and
\[
H(g_{X\to Y}(X,Z)|Z)\le I(X;Y)+\eta,
\]
where $\eta=\log(I(X;Y)+1)+4$. Consider the set
\[
A=\left\{ \left(H(g_{X\to Y}(X,z)),\,\E_{X}\left[d(X,\,g_{X\to Y}(X,z))\right]\right)\,:\,z\in\mathcal{Z}\right\} .
\]
\textcolor{black}{The point $(H(g_{X\to Y}(X,Z)|Z),\,\E\left[d(X,Y)\right])$ is a weighted
average of the points in $A$ (and thus is in the convex hull of $A$). Hence there exists $z$ satisfying the rate constraint $H(g_{X\to Y}(X,z)) \le H(g_{X\to Y}(X,Z)|Z)$, and there exists $z'$ satisfying the distortion constraint $\E_{X}\left[d(X,\,g_{X\to Y}(X,z'))\right] \le \E\left[d(X,Y)\right]$. However, there may not exist a single $z$ simultaneously satisfying both constraints. Hence we invoke
Carath\'eodory's theorem to find a mixture between two points $z_{0},z_{1}$ and $\lambda\in[0,1]$
such that both constraints are satisfied:}
\begin{align*}
(1-\lambda)H(g_{X\to Y}(X,z_{0}))+\lambda H(g_{X\to Y}(X,z_{1}))& \le H(g_{X\to Y}(X,Z)|Z)\le I(X;Y)+\eta,\\
(1-\lambda)\E_{X}\left[d(X,\,g_{X\to Y}(X,z_{0}))\right]+\lambda\E_{X}\left[d(X,\,g_{X\to Y}(X,z_{1}))\right]& \le\E\left[d(X,Y)\right].
\end{align*}
Note that to satisfy the above inequalities, we need one point less than stated in Carath\'eodory's theorem.
Take $Q\sim\Bern(\lambda)$, $\tilde{Y}=g_{X\to Y}(X,z_{Q})$.
Then
\textcolor{black}{
\[
H(\tilde{Y}) \le H(\tilde{Y}|Q)+H(Q) \le H(\tilde{Y}|Q)+1\le I(X;Y)+\eta+1.
\]
}
We use a Huffman code to encode $\tilde{Y}$ and obtain an expected length $\bar R \le H(\tilde{Y}) + 1$. The result follows by letting $\epsilon \to 0$.
\end{IEEEproof}
\medskip

\textcolor{black}{Note that a stochastic encoder is used in the proof. Nevertheless, the encoder only needs to randomize between two deterministic encoding functions in order to achieve Theorem~\ref{thm:oneshotlossy}.}

\textcolor{black}{An interesting implication of Theorem~\ref{thm:oneshotlossy} is that for any source $\P_X$, distortion measure $d(x,y)$, and distortion level $D$, the optimal asymptotic rate $R(D)$ cannot be too far from the optimal one-shot expected description length $\bar{R}^*(D)=\inf\{\bar{R}:\, (\bar{R},D)\; \text{achievable}\} \le R(D)+\log(R(D)+1)+6$. For example, there does not exist $(\P_X,d(x,y),D)$, where $R(D)=100$ but $\bar{R}^*(D)\ge 113$. This is a benefit of considering variable-length codes. Such conclusion does not hold if we consider fixed-length codes instead (e.g., if $X \sim \mathrm{Geom}(1/2)$, $d(x,y)=\mathbf{1}\{x \neq y\}$, then $R(D)\le 2$ for any $D \ge 0$, but the optimal length of the one-shot fixed-length code tends to infinity as $D\to 0$).
}

{\color{black}
Although the above achievability proof does not use random coding, it can be interpreted as using the following \emph{soft random coding} scheme.
% Operationally, a scheme for lossy source coding for discrete $Y$ based on exponential
%functional representation is described as follows.
%Comparisons with the conventional random codebook scheme are also made.
\begin{itemize}
\item[] \emph{Soft codebook generation}. The random variable $Z=\{(\tilde{Y}_{i}, T_{i})\}_{i=1,2,\ldots}$ produced by the Poisson functional representation represents the
choice of the codebook. 
We select a ``soft codebook'' by conditioning on $Z =\{(\tilde{y}_{i}, t_{i})\}_{i=1,2,\ldots}$.
Unlike conventional codebook $\mathcal{C}\subseteq Y$ which contains a fixed number of $y$'s, a soft codebook $\{(\tilde{y}_{i}, t_{i})\}$ contains an infinite sequence of $\tilde{y}_i$'s, each with a weight $t_i$ (the smaller $t_i$ is, the more likely $\tilde{y}_i$ is chosen).
\item[] \emph{Encoding}. The encoder observes $x$ and finds the reconstruction $\tilde{y}_k$ where
\[
k=\underset{i}{\arg\min}\,\,t_{i}\cdot\frac{d\,\P_{Y}}{d\,\P_{Y|X}(\cdot|x)}(\tilde{y}_{i}).
\]
It then encodes the index $k$ using an optimal prefix-free
code for the Zipf distribution~\eqref{eqn:zipf}. This is analogous to a conventional codebook generation in which we find the closest $y\in\mathcal{C}$ 
to $x$ and encodes it into its index in $\mathcal{C}$. Here
we use a prefix-free code over the positive integers to encode the index into the description $m$
because the index $k$ can be unbounded, but the smaller $k$'s (with smaller $t_k$'s)
are more likely to be used so they are assigned shorter descriptions.
\item[] \emph{Decoding}. The decoder receives $m$, recovers $k$, then outputs $\tilde{y}_k$.
\end{itemize}
}
\medskip

\textcolor{black}{Note that the soft random coding scheme shares some similarity with the likelihood encoder in~\cite{song2016likelihood}, which uses a conventional i.i.d. random codebook generation $y(m) \sim \P_{Y}$, $m = 1,\ldots,2^R$, but uses a stochastic encoder which chooses $m$ with probability proportional to the likelihood function
\[
\mathcal{L}(m|x) = p_{X|Y}(x|y(m)) \propto \frac{d \, \P_{Y|X}(\cdot|x) }{ d \, \P_Y}(y(m)).
\]
The soft random coding scheme can be viewed as fixing the randomness in the likelihood encoder as part of the codebook.
}

A related one-shot variable-length lossy source coding setting with a constraint on the probability that the distortion exceed certain level (instead of average distortion) was studied in~\cite{kostina2015variable}. In~\cite{posner1971epsilon}, a result similar to Theorem~\ref{thm:oneshotlossy} is given in the context of epsilon entropy.

The finite blocklength variable-length lossy source coding problem~\cite{pinkston1967encoding} concerns the case in which the source is memoryless and average per symbol distortion $d(x^n,y^n) = (1/n)\sum_i d(x_i,y_i)$. In~\cite{zhang1997redundancy} it is shown that the expected per symbol description length $\bar R / n = R(D) + (1+o(1)) (1/n)\log n$ is achievable via $d$-semifaithful codes~\cite{ornstein1990universal} with $d(X^n,\tilde{Y}^n) \le D$ surely. Applying Theorem~\ref{thm:oneshotlossy} to $X^n$, we have
\[
\bar R / n = R(D)+(1/n)(\log(n R(D)+1)+6) = R(D) + (1+o(1)) (1/n)\log n.
\]
Hence we achieve the same redundancy as~\cite{zhang1997redundancy} albeit under the expected distortion constraint instead of the stronger sure distortion constraint using the $d$-semifaithful codes.

We can use Theorem~\ref{thm:oneshotlossy} to establish the achievability of Shannon's (asymptotic) lossy source coding theorem~\cite{Shannon1959}, assuming there exists a symbol $y_0 \in \Yc$ with finite $d(x,y_0)$ for all $x$. First note that the redundancy $(1+o(1)) (1/n)\log n$ in the finite block length extension can be made arbitrarily small, hence $\bar R / n$ can be made arbitrarily close to $R(D)$. Now we use the finite block length scheme over $l$ blocks of $n$ source symbols each of length $n$ (for a total block length of $nl$). By the law of large numbers, the probability that the total description length is greater than $nl(R(D) + \epsilon)$ tends to $0$ as the block length approaches infinity. Hence, we can construct a fixed length code out of the variable-length code by simply discarding descriptions longer than $nl(R(D) + \epsilon)$ and assigning the reconstruction sequence $(y_0,\ldots,y_0)$ to the discarded descriptions.

\medskip

%I AM NOT SURE WE NEED THIS REMARK
%\begin{rem}
%Carath\'eodory's theorem implies that any point $x\in\mathbb{R}_{\ge0}^{n}$
%in the convex hull of a set $A\subseteq\mathbb{R}_{\ge0}^{n}$ can
%be expressed as a convex combination of $n+1$ points $\alpha_{1}v_{1}+\cdots+\alpha_{n+1}v_{n+1}=x$,
%$v_{1},\ldots,v_{n+1}\in A$. If we only require a point $\gamma x$
%for some $\gamma\in[0,1]$, then $\gamma x$ can be expressed as a
%convex combination of $n$ points. Suppose $\sum_{i=1}^{n+1}\beta_{i}v_{i}=0$
%(since they are linearly dependent), and without loss of generality
%assume $\sum_{i}\beta_{i}\ge0$. Note that at least one of $\beta_{i}$'s
%is negative. Then 
%\[
%\sum_{i}\frac{\alpha_{i}+\delta\beta_{i}}{\sum_{j}\left(\alpha_{j}+\delta\beta_{j}\right)}v_{i}=\frac{1}{\sum_{j}\left(\alpha_{j}+\delta\beta_{j}\right)}x,
%\]
%where $\delta\ge0$ is the smallest number such that at least one
%of $\alpha_{i}+\delta\beta_{i}$ is zero.
%\end{rem}

%\begin{rem}
%In ~\cite{winter2002compression}, Winter showed
%that Shannon's lossy source theorem~\cite{Shannon1959} follows
%from asymptotic channel simulation with sufficient common randomness in the asymptotic
%regime. The result in this section can be considered as a one-shot version of Winter's result.
%\end{rem}
%----------------------------------
\subsection{Multiple Description Coding
\label{sec:multidesc}}

In this section, we use the SFRL to establish a one-shot inner bound
for the variable-length multiple description coding problem, which
yields an alternative proof of the El Gamal-Cover inner bound~\cite{El-Gamal--Cover1982}
and the Zhang-Berger inner bound~\cite{Zhang--Berger1987,Venkataramani--Kramer--Goyal2003,Wang--Chen--Zhao--Cuff--Permuter2011}
in the asymptotic regime. The encoder observes $X\sim\P_{X}$ and
produces two prefix-free descriptions $M_{1},M_{2}\in\{0,1\}^{*}$.
Decoder 1 observes $M_{1}$ and generates $\tilde{Y}_{1}$ with distortion
$d_{1}(X,\tilde{Y}_{1})$. Similarly, Decoder 2 observes $M_{2}$
and produces $\tilde{Y}_{2}$ with distortion $d_{2}(X,\tilde{Y}_{2})$.
Decoder 0 observes $M_{1}$ and $M_{2}$ and produces $\tilde{Y}_{0}$
with distortion $d_{0}(X,\tilde{Y}_{0})$. An expected description
length-distortion tuple $(\bar{R}_{1},\bar{R}_{2},D_{0},D_{1},D_{2})$
is said to be achievable if there exists a scheme with expected description
length $\E[L(M_{i})]\le\bar{R}_{i}$ and expected distortion $\E[d_{i}(X,\tilde{Y}_{i})]\le D_{i}$.
\begin{thm}
\label{thm:multidesc}The tuple $(\bar{R}_{1},\bar{R}_{2},D_{0},D_{1},D_{2})$
is achievable if 
\begin{align*}
\bar{R}_{1} & \ge I(X;Y_{1},U)+2\eta,\\
\bar{R}_{2} & \ge I(X;Y_{2},U)+2\eta,\\
\bar{R}_{1}+\bar{R}_{2} & \ge I(X;Y_{0},Y_{1},Y_{2}|U)+2I(X;U)+I(Y_{1};Y_{2}|U)+5\eta,\\
D_{i} & \ge\E[d_{i}(X,Y_{i})]\;\;\text{for \ensuremath{i=0,1,2}}
\end{align*}
for some $\P_{U,Y_{0},Y_{1},Y_{2}|X}$, where
\[
\eta=\log\big(I(X;Y_{0},Y_{1},Y_{2},U)+I(Y_{1};Y_{2}|U)+1\big)+7.
\]
\end{thm}
Note that the only difference between the above region and Zhang-Berger
inner bound is the addition of $\eta$, which grows like $\log n$
if we consider $X^{n}$ and does not affect the asymptotic rate.
\begin{IEEEproof}
It suffices to prove the achievability of the corner point: 
\begin{align}
\bar{R}_{1} & =I(X;Y_{1}|U)+I(X;U)+2\eta-1, \label{eqn:mdc_alt1}\\
\bar{R}_{2} & =I(X,Y_{1};Y_{2}|U)+I(X;Y_{0}|Y_{1},Y_{2},U)+I(X;U)+3\eta-1, \label{eqn:mdc_alt2}\\
D_{i} & =\E[d_{i}(X,Y_{i})]\;\;\text{for \ensuremath{i=0,1,2}}. \label{eqn:mdc_alt3}
\end{align}
The desired rate region can be achieved by time sharing between this
corner point and the other corner point where $Y_{1},Y_{2}$ are flipped,
resulting in a penalty of at most $1$ bit (we can use the first bits
of $M_{1}$ and $M_{2}$ to represent which corner point it is).

Applying the SFRL~\eqref{eqn:sfrl_hyz} to $X,U$, we have $U=g_{X\to U}(X,Z_{3})$, where
$Z_{3}\indep X$ such that 
\begin{align*}
H(U|Z_{3}) & \le I(X;U)+\log(I(X;U)+1)+4\\
 & \le I(X;U)+\eta-3.
\end{align*}
Applying the SFRL to $X,Y_{1}$ conditioned on $U$~\eqref{eqn:sfrl_cond}, we have $Y_{1}=g_{X\to Y_{1}|U}(X,Z_{1},U)$,
where $Z_{1}\indep(X,U)$ such that 
\begin{align*}
H(Y_{1}|U,Z_{1}) & \le I(X;Y_{1}|U)+\log(I(X;Y_{1}|U)+1)+4\\
 & \le I(X;Y_{1}|U)+\eta-3.
\end{align*}
Applying the SFRL to $(X,Y_{1}),Y_{2}$ conditioned on $U$, we have
$Y_{2}=g_{XY_{1}\to Y_{2}|U}(X,Y_{1},Z_{2},U)$, $Z_{2}\indep(X,Y_{1},U)$
such that 
\begin{align*}
H(Y_{2}|U,Z_{2}) & \le I(X,Y_{1};Y_{2}|U)+\log(I(X,Y_{1};Y_{2}|U)+1)+4\\
 & \le I(X,Y_{1};Y_{2}|U)+\eta-3.
\end{align*}
Applying the SFRL to $X,Y_{0}$ conditioned on $(Y_{1},Y_{2},U)$,
we have $Y_{0}=g_{X\to Y_{0}|Y_{1}Y_{2}U}(X,Z_{0},Y_{1},Y_{2},U)$,
$Z_{0}\indep(X,Y_{1},Y_{2},U)$ such that 
\begin{align*}
H(Y_{0}|Y_{1},Y_{2},U,Z_{0}) & \le I(X;Y_{0}|Y_{1},Y_{2},U)+\log(I(X;Y_{0}|Y_{1},Y_{2},U)+1)+4\\
 & \le I(X;Y_{0}|Y_{1},Y_{2},U)+\eta-3.
\end{align*}
Note that $Z_{0}^{3}\indep X$. Consider the convex hull
of the 7-dimensional vectors 
\[
\left[\begin{array}{c}
H(U|Z_{0}^{3}=z_{0}^{3})\\
H(Y_{1}|U,Z_{0}^{3}=z_{0}^{3})\\
H(Y_{2}|U,Z_{0}^{3}=z_{0}^{3})\\
H(Y_{0}|Y_{1},Y_{2},U,Z_{0}^{3}=z_{0}^{3})\\
\E[d_{0}(X,Y_{0})\,|\,Z_{0}^{3}=z_{0}^{3}]\\
\E[d_{1}(X,Y_{1})\,|\,Z_{0}^{3}=z_{0}^{3}]\\
\E[d_{2}(X,Y_{2})\,|\,Z_{0}^{3}=z_{0}^{3}]
\end{array}\right]
\]
for different $z_{0}^{3}\in\mathcal{Z}_{0}\times\mathcal{Z}_{1}\times\mathcal{Z}_{2}\times\mathcal{Z}_{3}$.
By Carath\'eodory's theorem, there exists a pmf $p_{Q}$ with cardinality
$\left|\mathcal{Q}\right|\le7$ and $\tilde{z}_{0}^{3}(q)$ such that
\[
H(U|Q,\,Z_{0}^{3}=\tilde{z}_{0}^{3}(Q))\le I(X;U)+\eta-3,
\]
and similarly for the other 6 inequalities. Take $\tilde{U}=g_{X\to U}(X,\,\tilde{z}_{3}(Q))$,
$\tilde{Y}_{1}=g_{X\to Y_{1}|U}(X,\,\tilde{z}_{1}(Q),\tilde{U})$,
$\tilde{Y}_{2}=g_{XY_{1}\to Y_{2}|U}(X,\tilde{Y}_{1},\allowbreak\tilde{z}_{2}(Q),\tilde{U})$
and $\tilde{Y}_{0}=g_{X\to Y_{0}|Y_{1}Y_{2}U}(X,\allowbreak\tilde{z}_{0}(Q),\tilde{Y}_{1},\tilde{Y}_{2},\tilde{U})$.
Write $C_{p_{Y}}(y)\in\{0,1\}^{*}$ for the Huffman codeword of $y$
for the distribution $p_{Y}$. We set $M_{1}$ to be the concatenation
of $Q$ (3 bits), $C_{p_{\tilde{U}|Q}(\,\cdot\,|Q)}(\tilde{U})$ and
$C_{p_{\tilde{Y}_{1}|\tilde{U}Q}(\,\cdot\,|\tilde{U},Q)}(\tilde{Y}_{1})$,
and $M_{2}$ to be the concatenation of $Q$, $C_{p_{\tilde{U}|Q}(\,\cdot\,|Q)}(\tilde{U})$,
$C_{p_{\tilde{Y}_{2}|\tilde{U}Q}(\,\cdot\,|\tilde{U},Q)}(\tilde{Y}_{2})$
and $C_{p_{\tilde{Y}_{0}|\tilde{Y}_{1}\tilde{Y}_{2}\tilde{U}Q}(\,\cdot\,|\tilde{Y}_{1},\tilde{Y}_{2},\tilde{U},Q)}(\tilde{Y}_{0})$.
The expected length of $M_{1}$ is upper bounded by
\begin{align*}
 & 3+\left(I(X;U)+\eta-3+1\right)+\left(I(X;Y_{1}|U)+\eta-3+1\right)\\
 & =I(X;Y_{1}|U)+I(X;U)+2\eta-1.
\end{align*}
\textcolor{black}{Hence~\eqref{eqn:mdc_alt1} is satisfied. By similar arguments,~\eqref{eqn:mdc_alt2} and~\eqref{eqn:mdc_alt3} hold.}
%The bound on the expected length of $M_{2}$ can be obtained similarly.

Decoder 1 receives $M_{1}$ and recovers $Q$, and then recovers $\tilde{U}$
by decoding the Huffman code for the distribution $p_{\tilde{U}|Q}(\,\cdot\,|Q)$,
and then recovers $\tilde{Y}_{1}$ similarly. Decoder 2 receives $M_{2}$
and recovers $Q,\tilde{U}$ and $\tilde{Y}_{2}$. Decoder 0 receives
$M_{1},M_{2}$ and recovers $Q,\tilde{U},\tilde{Y}_{1},\tilde{Y}_{2}$
and $\tilde{Y}_{0}$. 
\end{IEEEproof}

\subsection{Lossy Gray--Wyner System \label{sec:gw}}

In this section, we use the SFRL to establish a one-shot
inner bound for the lossy Gray--Wyner system~\cite{Gray--Wyner1974},
which yields an alternative proof of the achievability of the rate region in the asymptotic
regime. The encoder observes $(X_{1},X_{2})\sim\P_{X_{1},X_{2}}$
and produces three prefix-free descriptions $M_{0},M_{1},M_{2}\in\{0,1\}^{*}$.
Decoder 1 observes $M_{0},M_{1}$ and generates $\tilde{Y}_{1}$ with
distortion $d_{1}(X_{1},\tilde{Y}_{1})$. Similarly, Decoder 2 observes
$M_{0},M_{2}$ and produces $\tilde{Y}_{2}$ with distortion $d_{2}(X_{2},\tilde{Y}_{2})$.
An expected description length-distortion tuple $(\bar{R}_{0},\bar{R}_{1},\bar{R}_{2},D_{1},D_{2})$
is said to be achievable if there exists a scheme with expected description
length $\E[L(M_{i})]\le\bar{R}_{i}$ and expected distortion $\E[d_{i}(X_{i},\tilde{Y}_{i})]\le D_{i}$.
\begin{thm}
\label{thm:multidesc-1}The tuple $(\bar{R}_{0},\bar{R}_{1},\bar{R}_{2},D_{1},D_{2})$
is achievable if 
\begin{align}
\bar{R}_{0} & \ge I(X_{1},X_{2};U)+\log(I(X_{1},X_{2};U)+1)+8, \label{eqn:gws1}\\
\bar{R}_{1} & \ge I(X_{1};Y_{1}|U)+\log(I(X_{1};Y_{1}|U)+1)+5, \label{eqn:gws2}\\
\bar{R}_{2} & \ge I(X_{2};Y_{2}|U)+\log(I(X_{2};Y_{2}|U)+1)+5, \label{eqn:gws3}\\
D_{i} & \ge\E[d_{i}(X_{i},Y_{i})]\;\;\text{for \ensuremath{i=1,2}} \label{eqn:gws4}
\end{align}
for some $\P_{U|X_{1},X_{2}}$, $\P_{Y_{1}|X_{1},U}$, $\P_{Y_{2}|X_{2},U}$.
\end{thm}
Note that the only difference between the above region
and the lossy Gray--Wyner rate region~\cite[p. 357]{elgamal2011network} is the addition of the logarithm terms,
which grows like $\log n$ if we consider $X_{1}^{n},X_{2}^{n}$ and
does not affect the asymptotic rate.
\begin{IEEEproof}
Applying the SFRL to $(X_{1},X_{2}),U$, we have
$U=g_{X_{1}X_{2}\to U}(X_{1},X_{2},Z_{0})$, where $Z_{0}\indep(X_{1},X_{2})$
such that 
\begin{align*}
H(U|Z_{0}) & \le I(X_{1},X_{2};U)+\log(I(X_{1},X_{2};U)+1)+4.
\end{align*}
Applying the SFRL to $X_{1},Y_{1}$ conditioned on $U$~\eqref{eqn:sfrl_cond}, we have $Y_{1}=g_{X_{1}\to Y_{1}|U}(X_{1},Z_{1},U)$,
where $Z_{1}\indep(X_{1},U)$ such that 
\begin{align*}
H(Y_{1}|U,Z_{1}) & \le I(X_{1};Y_{1}|U)+\log(I(X_{1};Y_{1}|U)+1)+4.
\end{align*}
Applying the SFRL to $X_{2},Y_{2}$ conditioned on $U$, we have $Y_{2}=g_{X_{2}\to Y_{2}|U}(X_{2},Z_{2},U)$,
where $Z_{2}\indep(X_{2},U)$ such that 
\begin{align*}
H(Y_{2}|U,Z_{2}) & \le I(X_{2};Y_{2}|U)+\log(I(X_{2};Y_{2}|U)+1)+4.
\end{align*}
Note that $Z_{0}^{2}\indep(X_{1},X_{2})$. Consider the
convex hull of the 5-dimensional vectors 
\[
\left[\begin{array}{c}
H(U|Z_{0}^{2}=z_{0}^{2})\\
H(Y_{1}|U,Z_{0}^{2}=z_{0}^{2})\\
H(Y_{2}|U,Z_{0}^{2}=z_{0}^{2})\\
\E[d_{1}(X_{1},Y_{1})\,|\,Z_{0}^{2}=z_{0}^{2}]\\
\E[d_{2}(X_{2},Y_{2})\,|\,Z_{0}^{2}=z_{0}^{2}]
\end{array}\right]
\]
for different $z_{0}^{2}\in\mathcal{Z}_{0}\times\mathcal{Z}_{1}\times\mathcal{Z}_{2}$.
By Carath\'eodory's theorem, there exists a pmf $p_{Q}$ with cardinality
$\left|\mathcal{Q}\right|\le5$ and $\tilde{z}_{0}^{2}(q)$ such that
\[
H(U|Q,\,Z_{0}^{2}=\tilde{z}_{0}^{2}(Q))\le I(X_{1},X_{2};U)+\log(I(X_{1},X_{2};U)+1)+4,
\]
and similarly for the other 4 inequalities. Take $\tilde{U}=g_{X_{1}X_{2}\to U}(X_{1},X_{2},\,\tilde{z}_{0}(Q))$,
$\tilde{Y}_{1}=g_{X_{1}\to Y_{1}|U}(X_{1},\,\tilde{z}_{1}(Q),\tilde{U})$
and $\tilde{Y}_{2}=g_{X_{2}\to Y_{2}|U}(X_{2},\,\tilde{z}_{2}(Q),\tilde{U})$.
Write $C_{p_{Y}}(y)\in\{0,1\}^{*}$ for the Huffman codeword of $y$
for the distribution $p_{Y}$. We set $M_{0}$ to be the concatenation
of $Q$ (3 bits) and $C_{p_{\tilde{U}|Q}(\,\cdot\,|Q)}(\tilde{U})$,
$M_{1}=C_{p_{\tilde{Y}_{1}|\tilde{U}Q}(\,\cdot\,|\tilde{U},Q)}(\tilde{Y}_{1})$
and $M_{2}=C_{p_{\tilde{Y}_{2}|\tilde{U}Q}(\,\cdot\,|\tilde{U},Q)}(\tilde{Y}_{2})$.
The expected length of $M_{0}$ is upper bounded by
\begin{align*}
 & 3+\left(H(U|Z_{0})+1\right)\\
 & \le3+\left(I(X_{1},X_{2};U)+\log(I(X_{1},X_{2};U)+1)+4+1\right)\\
 & =I(X_{1},X_{2};U)+\log(I(X_{1},X_{2};U)+1)+8.
\end{align*}
\textcolor{black}{Hence~\eqref{eqn:gws1} is satisfied. By similar arguments,~\eqref{eqn:gws2},~\eqref{eqn:gws3} and~\eqref{eqn:gws4} hold.}
% The bound on the expected length of $M_{1},M_{2}$ can be obtained similarly.

Decoder 1 receives $M_{0},M_{1}$ and recovers $Q$,
and then recovers $\tilde{U}$ by decoding the Huffman code for the
distribution $p_{\tilde{U}|Q}(\,\cdot\,|Q)$, and then recovers $\tilde{Y}_{1}$
by decoding the Huffman code for the distribution $p_{\tilde{Y}_{1}|\tilde{U}Q}(\,\cdot\,|\tilde{U},Q)$.
Similar for Decoder 2.
\end{IEEEproof}

%---------------------------------------------
\section{Achievability of Gelfand--Pinsker \label{sec:state}}

%\subsection{Gelfand--Pinsker\label{sec:distlossy}}
In this section, we use  the SFRL to prove the achievability part
of the Gelfand-Pinsker theorem~\cite{Gelfand--Pinsker1980a} for discrete memoryless channels
with discrete memoryless state $p_{S} p_{Y|X,S}$, where the state is noncausally available at the encoder. The asymptotic capacity of this setting is
\[
C_\mathrm{GP}= \max_{p_{U|S},\,x(u,s)}\left(I(U;Y)-I(U;S)\right).
\]
We show the achievability of any rate below $C_\mathrm{GP}$ directly by using the SFRL to reduce the channel to a point-to-point memoryless channel. Fix $p_{U|S}$ and $x(u,s)$ that attain the capacity. Applying the SFRL to $S,U$, there exists a random variable $V\indep S$
such that
\[
H(U|V)\le I(U;S)+\log(I(U;S)+1)+4.
\]
Note that
\begin{align*}
I(V;Y) & =I(U;Y)-I(U;Y|V)+I(V;Y|U)\\
 & \ge I(U;Y)-H(U|V)\\
 & \ge I(U;Y)-I(U;S)-\log(I(U;S)+1)-4.
\end{align*}
Hence we have constructed a memoryless point-to-point channel $p_{Y|V}$ with achievable rate close to $I(U;Y)-I(U;S)$.

For $n$ channel uses, let $U^{n}|\{S^{n}=s^{n}\}\sim\prod_{i}p_{U|S}(u_{i}|s_{i})$. The
SFRL applied to $S^{n},U^{n}$ gives
\[
I(V;Y^{n})\ge nI(U;Y)-nI(U;S)-\log(nI(U;S)+1)-4.
\]
Now we use the channel $p_{Y^{n}|V}$ $l$ times (for a total block length of
$nl$). By the channel coding theorem, we can communicate
$l(nI(U;Y)-nI(U;S)-\log(nI(U;S)+1)-4)-o(l)$ bits with error probability that
tends to 0 as $l\to\infty$. Letting $n\to\infty$ completes the proof.

In the above proof, we see that the SFRL can be used to convert a channel with state
into a point-to-point channel by ``orthogonalizing'' the
auxiliary input $U$ and the state $S$. The point-to-point
channel can be constructed explicitly via Poisson functional representation.
This construction can be useful for designing codes for channels with
state based on codes for point-to-point channels. It is interesting to note that this reduction makes the achievability proof for the Gelfand--Pinsker quite similar to that for the causal case in which the channel is reduced to a point-to-point channel using the "Shannon strategy'' (see~\cite[p. 176]{elgamal2011network}).

Note that Marton's inner bound for the broadcast channels with private
messages~\cite{Marton1979} can also be proved using the SFRL in a similar manner. The idea is to ``orthogonalize'' the dependent auxiliary
random variables $U_{1},U_{2}$ by applying the SFRL on $U_1,U_2$
to produce two independent input random variables, and treat them
with $Y_{1},Y_{2}$ as an interference channel, and finally to treat
interference as noise.

\section{\textcolor{black}{Lower bound and properties of $I(X;Z|Y)$}\label{sec:tightness}}

%The SFRL states that $I(X;Z|Y)\le\log(I(X;Y)+1)+4$ is achievable. A natural
%question to ask is whether this inequality is tight for some $X,Y$. In this section, we
%show that the log term is necessary and that the SFRL is in general tight to within 5 bits.

Define the \emph{excess functional information} as
\[
\Psi(X\to Y)=\inf_{Z:\,Z\indep X,\,H(Y|X,Z)=0}I(X;Z|Y).
\]
\textcolor{black}{An equivalent way to state SFRL is  $\Psi(X\to Y)\le\log(I(X;Y)+1)+4$. In this section, we explore the properties of $\Psi(X\to Y)$. We first establish a lower bound.}
\begin{prop}
\label{prop:phi_lb}For discrete $Y$,
\[
\Psi(X\to Y)\ge-\sum_{y\in\mathcal{Y}}\int_{0}^{1}\P_{X}\left\{ p_{Y|X}(y|X)\ge t\right\} \log\left(\P_{X}\left\{ p_{Y|X}(y|X)\ge t\right\} \right)dt-I(X;Y).
\]
Moreover for $|\mathcal{Y}|=2$, equality holds in the above
inequality, and the infimum in $\Psi(X\to Y)$ is attained via the Poisson
functional representation.
\end{prop}
\begin{IEEEproof}
Fix $Z\indep X$ such that $Y=g(X,Z)$. For any $y$,
let $V_{y}=p_{Y|Z}(y|Z)$, $U\sim\U[0,1]$, $\tilde{X}_{y}=p_{Y|X}(y|X)$,
$\tilde{V}_{y}=\P\left\{ \tilde{X}_{y}\ge U\,|\,U\right\} $, then
$\E[V_{y}]=\E[\tilde{V}_{y}]=p_{Y}(y)$. We have{\allowdisplaybreaks
\begin{align*} 
 \int_{v}^{1}\P\{V_{y}\ge t\}dt
 & =\E\left[\max\left\{ V_{y}-v,\,0\right\} \right]\\
 & =\E_{Z}\left[\max\left\{ p_{Y|Z}(y|Z)-v,\,0\right\} \right]\\
 & =\E_{Z}\left[\max\left\{ \P_{X}\left\{ g(X,Z)=y\,|\,Z\right\} -v,\,0\right\} \right]\\
 & =\E_{Z}\left[\max\left\{ \E_{\tilde{X}_{y}}\left[\P_{X}\left\{ g(X,Z)=y\,\big|\,Z,\tilde{X}_{y}\right\} \,\Big|\,Z\right]-v,\,0\right\} \right]\\
 & =\E_{Z}\left[\max\left\{ \E_{\tilde{X}_{y}}\left[\P_{X}\left\{ g(X,Z)=y\,\big|\,Z,\tilde{X}_{y}\right\} \,\Big|\,Z\right]-\E_{\tilde{X}_{y}}\left[\mathbf{1}\left\{ \tilde{X}_{y}>F_{\tilde{X}_{y}}^{-1}(1-v)\right\} \right],\,0\right\} \right]\\
 & \le\E_{Z}\left[\E_{\tilde{X}_{y}}\left[\max\left\{ \P_{X}\left\{ g(X,Z)=y\,\big|\,Z,\tilde{X}_{y}\right\} -\mathbf{1}\left\{ \tilde{X}_{y}>F_{\tilde{X}_{y}}^{-1}(1-v)\right\} \,,\,0\right\} \,\Big|\,Z\right]\right]\\
 & =\E_{Z}\left[\E_{\tilde{X}_{y}}\left[\P_{X}\left\{ g(X,Z)=y\,\big|\,Z,\tilde{X}_{y}\right\} \cdot\mathbf{1}\left\{ \tilde{X}_{y}\le F_{\tilde{X}_{y}}^{-1}(1-v)\right\} \,\Big|\,Z\right]\right]\\
 & =\E_{\tilde{X}_{y}}\left[\E_{Z}\left[\P_{X}\left\{ g(X,Z)=y\,\big|\,Z,\tilde{X}_{y}\right\} \,\Big|\,\tilde{X}_{y}\right]\cdot\mathbf{1}\left\{ \tilde{X}_{y}\le F_{\tilde{X}_{y}}^{-1}(1-v)\right\} \right]\\
 & =\E_{\tilde{X}_{y}}\left[\E_{X}\left[\P_{Z}\left\{ g(X,Z)=y\,|\,X\right\} \,\big|\,\tilde{X}_{y}\right]\cdot\mathbf{1}\left\{ \tilde{X}_{y}\le F_{\tilde{X}_{y}}^{-1}(1-v)\right\} \right]\\
 & =\E_{\tilde{X}_{y}}\left[\E_{X}\left[p_{Y|X}(y|X)\,\big|\,\tilde{X}_{y}\right]\cdot\mathbf{1}\left\{ \tilde{X}_{y}\le F_{\tilde{X}_{y}}^{-1}(1-v)\right\} \right]\\
 & =\E_{\tilde{X}_{y}}\left[\tilde{X}_{y}\cdot\mathbf{1}\left\{ \tilde{X}_{y}\le F_{\tilde{X}_{y}}^{-1}(1-v)\right\} \right]\\
 & =\E_{U}\left[\max\left\{ \P\left\{ \tilde{X}_{y}\ge U\,\big|\,U\right\} -v,\,0\right\} \right]\\
 & =\E\left[\max\left\{ \tilde{V}_{y}-v,\,0\right\} \right]\\
 & =\int_{v}^{1}\P\{\tilde{V}_{y}\ge t\}dt.
\end{align*}}
Hence $V_{y}$ dominates $\tilde{V}_{y}$ stochastically in the second
order. By the concavity of $-t\log t$, we have
\begin{equation}
\begin{aligned}H(Y|Z) & =-\sum_{y}\E_{Z}\left[p_{Y|Z}(y|Z)\log p_{Y|Z}(y|Z)\right]\\
 & =-\sum_{y}\E\left[V_{y}\log V_{y}\right]\\
 & \ge-\sum_{y}\E\left[\tilde{V}_{y}\log\tilde{V}_{y}\right]\\
 & =-\sum_{y}\int_{0}^{1}\P_{X}\left\{ p_{Y|X}(y|X)\ge u\right\} \log\left(\P_{X}\left\{ p_{Y|X}(y|X)\ge u\right\} \right)du.
\end{aligned}
\label{eq:psi_hyz}
\end{equation}
Therefore,
\[
I(X;Z|Y)\ge-\sum_{y}\int_{0}^{1}\P_{X}\left\{ p_{Y|X}(y|X)\ge t\right\} \log\left(\P_{X}\left\{ p_{Y|X}(y|X)\ge t\right\} \right)dt-I(X;Y).
\]
One can verify that for $|\mathcal{Y}|=2$, equality in \eqref{eq:psi_hyz} holds by the definition of Poisson functional representation.
\end{IEEEproof}
The following proposition shows that there exists a sequence of $(X,Y)$ for which the bound $\Psi(X,Y)\le\log(I(X;Y)+1)+4$
given in the SFRL is tight within 5 bits. \textcolor{black}{An example where the $\log$ term is tight is also given in~\cite{braverman2014public}, though the additive constant is not specified there.}
\begin{prop} \label{prop:lb_example}
For every $\alpha \ge 0$, there exists discrete $X,Y$ such that $I(X;Y) \ge \alpha$ and
\[
\Psi(X\to Y)\ge\log(I(X;Y)+1)-1.
\]
\end{prop}

The proof is given in Appendix~\ref{subsec:lb_example}. Besides the upper bound given by the SFRL and its tightness, in the following we establish other
properties of $\Psi(X\to Y)$. We write the conditional
excess functional information as
\[
\Psi(X\to Y\,|\,Q)=\E_{Q}\left[\Psi(X\to Y\,|\,Q=q)\right].
\]

\begin{prop}
The excess functional information $\Psi(X\to Y)$ satisfies the following properties.

\begin{enumerate}
\item Alternative characterization.
\[
\Psi(X\to Y)=\inf_{Z:\,Z\indep X}H(Y|Z)-I(X;Y).
\]
\item Monotonicity. If $X_{1}\indep X_{2}$ and $X_{1}\indep(X_{2},Y_{2})\,|\,Y_{1}$,
then 
\[
\Psi((X_{1},X_{2})\to(Y_{1},Y_{2}))\ge\Psi(X_{1}\to Y_{1}).
\]
\item Subadditivity. If $(X_{1},Y_{1})\indep(X_{2},Y_{2})$,
then 
\[
\Psi((X_{1},X_{2})\to(Y_{1},Y_{2}))\le\Psi(X_{1}\to Y_{1})+\Psi(X_{2}\to Y_{2}).
\]
As a result, if we further have $X_{2}\indep Y_{2}$, then
$\Psi((X_{1},X_{2})\to(Y_{1},Y_{2}))=\Psi(X_{1}\to Y_{1})$ by monotonicity.
\item Data processing of $\Psi+I$.
%Write $(\Psi+I)(X\to Y)=\Psi(X\to Y)+I(X;Y)$ (we call this total functional information), then
If $X_{2}-X_{1}-Y_{1}-Y_{2}$
forms a Markov chain,
\[
\Psi(X_{1}\to Y_{1}) + I(X_1;Y_1)\ge\Psi(X_{2}\to Y_{2}) + I(X_2;Y_2).
\]
\item Upper bound by common entropy.
\[
\Psi(X\to Y)\le G(X;Y)-I(X;Y)\le\min\left\{ H(X|Y),\,H(Y|X)\right\} ,
\]
where $G(X;Y)=\min_{X\indep Y|W}H(W)$ is the common entropy
\cite{kumar2014exact,distsimcont}.

\item Conditioning. If $Q$ satisfies $H(Q|X)=0$, then
\[
\Psi(X\to Y)\ge\Psi(X\to Y\,|\,Q).
\]
If we further have $H(Q|Y)=0$, then equality holds in the above inequality.
\item Successive minimization.
\[
\Psi(X\to Y)=\inf_{V:\,V\indep X}\left\{ I(X;V|Y)+\Psi(X\to Y\,|\,V)\right\} .
\]
\end{enumerate}
\end{prop}
\begin{IEEEproof}
\begin{enumerate}
\item \emph{Alternative characterization}. Note that if $Z\indep X$
and $H(Y|X,Z)=0$, then $H(Y|Z)-I(X;Y)=I(X;Z|Y)$, hence
\[
\inf_{Z:\,Z\indep X,\,H(Y|X,Z)=0}I(X;Z|Y)\ge\inf_{Z:\,Z\indep X}H(Y|Z)-I(X;Y).
\]
For the other direction, assume $Z\indep X$. By the functional
representation lemma, let $Y=g(X,Z,\tilde{Z})$, $\tilde{Z}\indep(X,Z)$.
We have
\begin{align*}
H(Y|Z)-I(X;Y) & \ge H(Y|Z,\tilde{Z})-I(X;Y)\\
 & =I(X;Z,\tilde{Z}|Y)\\
 & \ge\inf_{Z':\,Z'\indep X,\,H(Y|X,Z')=0}I(X;Z'|Y).
\end{align*}

%\[
%\begin{aligned}I(X;Z|Y) & \le I(X;Z|Y)+H(X|Y,Z)=H(X|Y),\end{aligned}
%\]
%\[
%\begin{aligned}I(X;Z|Y) & =I(X;Z)+I(Y;Z|X)-I(Y;Z)\\
% & \le I(Y;Z|X)\\
% & \le I(Y;Z|X)+H(Y|X,Z)\\
% & =H(Y|X).
%\end{aligned}
%\]

\item \emph{Monotonicity}. Let $Z$ satisfies $Z\indep(X_{1},X_{2})$
and $H(Y_{1},Y_{2}|X_{1},X_{2},Z)=0$. Note that $(Z,X_{2})\indep X_{1}$
and $H(Y_{1}|X_{1},Z,X_{2})=0$. Hence
\begin{align*}
I(X_{1},X_{2};Z|Y_{1},Y_{2}) & \ge I(X_{1};Z|X_{2},Y_{1},Y_{2})\\
 & =I(X_{1};Z|X_{2},Y_{1},Y_{2})+I(X_{1};Y_{2}|X_{2},Y_{1})\\
 & =I(X_{1};Z|X_{2},Y_{1})+I(X_{1};Y_{2}|X_{2},Y_{1},Z)\\
 & \ge I(X_{1};Z|X_{2},Y_{1})\\
 & =I(X_{1};Z|X_{2},Y_{1})+I(X_{1};X_{2}|Y_{1})\\
 & =I(X_{1};Z,X_{2}|Y_{1})\\
 & \ge\Psi(X_{1}\to Y_{1}).
\end{align*}

\item \emph{Subadditivity}. let $Z_{1},Z_{2}$ satisfies $Z_{i}\indep X_{i}$
and $H(Y_{i}|X_{i},Z_{i})=0$, then
\begin{align*}
\Psi((X_{1},X_{2})\to(Y_{1},Y_{2})) & \le I(X_{1},X_{2};\,Z_{1},Z_{2}\,|\,Y_{1},Y_{2})\\
 & =I(X_{1};Z_{1}|Y_{1})+I(X_{2};Z_{2}|Y_{2}).
\end{align*}

\item \emph{Data processing of $\Psi+I$}. let $Z\indep X_1$, and let $Y_2=g(Y_1,W)$ be the functional representation of $Y_2$. Then $(Z,W)\indep X_2$, and by the the alternative characterization,
\begin{align*}
\Psi(X_{2}\to Y_{2}) + I(X_2;Y_2) & \le H(Y_2 | Z,W)\\
 & =H(Y_2 | Z,W,Y_1) + I(Y_1;Y_2 | Z,W)\\
 & \le H(Y_1 | Z,W)\\
 & = H(Y_1 | Z).
\end{align*}

\item The upper bound by common entropy is a direct consequence of the data processing inequality in the previous part.

\item \emph{Conditioning}. Assume that $H(Q|X)=0$, $Z\indep X$
and $H(Y|X,Z)=0$, then $Z\indep X|\{Q=q\}$and $H(Y|X,Z,Q=q)=0$
for all $q$, hence
\begin{align*}
I(X;Z|Y) & \ge I(X;Z|Y,Q)\\
 & =\E_{q\sim\mathrm{P}_{Q}}\left[I(X;Z|Y,Q=q)\right]\\
 & \ge\E_{q\sim\mathrm{P}_{Q}}\left[\Psi(X\to Y\,|\,Q=q)\right].
\end{align*}
To show the equality case, assume $H(Q|Y)=0$. Let $\tilde{Z}$ satisfies
$\tilde{Z}\indep X|\{Q=q\}$ and $H(Y|X,\tilde{Z},Q=q)=0$
for all $q$. By functional representation lemma, let $\tilde{Z}=g(Q,Z)$,
$Z\indep Q$, and since we are invoking functional representation
lemma over the marginal distribution of $(Q,\tilde{Z})$, we can assume
$Z\indep(X,Y)|(Q,\tilde{Z})$. Hence $Z\indep X$.
We have
\begin{align*}
\E_{q\sim\mathrm{P}_{Q}}\left[I(X;\tilde{Z}|Y,Q=q)\right] & =I(X;\tilde{Z}|Y,Q)\\
 & =I(X;Z|Y,Q)\\
 & =I(X;Z|Y)\\
 & \ge\Psi(X\to Y).
\end{align*}

\item \emph{Successive minimization}. Assume that $V\indep X$, and
let $\tilde{Z}$ satisfy $\tilde{Z}\indep X|\{V=v\}$
and $H(Y|X,\tilde{Z},V=v)=0$ for all $v$, then $X\indep(\tilde{Z},V)$. We have
\begin{align*}
\E_{q\sim\P_{Q}}\left[I(X;\tilde{Z}|Y,V=v)\right] & =I(X;\tilde{Z}|Y,V)\\
 & =I(X;\tilde{Z},V|Y)-I(X;V|Y)\\
 & =I(X;\tilde{Z},V|Y)-I(X;V|Y)\\
 & \ge\Psi(X\to Y)-I(X;V|Y).
\end{align*}
Note that $I(X;V|Y)+\Psi(X\to Y\,|\,V)=\Psi(X\to Y)$ if $V=\emptyset$.
Also note that
\begin{align*} 
 \inf_{V:\,V\indep X}\left\{ I(X;V|Y)+\Psi(X\to Y\,|\,V)\right\} 
 & \le\inf_{V:\,V\indep X,\,H(Y|X,Z)=0}\left\{ I(X;V|Y)+\Psi(X\to Y\,|\,V)\right\} \\
 & =\inf_{V:\,V\indep X,\,H(Y|X,Z)=0}I(X;V|Y)\\
 & =\Psi(X\to Y).
\end{align*}
\end{enumerate}
\end{IEEEproof}
\begin{rem}
If $\Psi(X,Y)=0$, then it means that there exists $Z$ such that
$Z\indep X$, $Z\indep X|Y$, $H(Y|Z)=I(X;Y)$
and $H(Y|X,Z)=0$. This implies there exists $z$ such that $H(Y|Z=z)\ge I(X;Y)$
and $H(Y|X,Z=z)=0$. Hence it is possible to perform one-shot zero
error channel coding on the channel $\P_{X|Y}$ with input distribution
$\P_{Y|Z=z}$ to communicate a message with entropy $\ge I(X;Y)$.
%The upper bound $\Psi(X\to Y)\le\min\left\{ H(X|Y),\,H(Y|X)\right\} $,
%the subadditivity property, grouping property and successive minimization
%property are useful for proving whether $\Psi(X,Y)=0$.
\end{rem}

%\begin{rem}
%Consider i.i.d. $(X_{i},Y_{i})$. We have $\Psi(X_{1},Y_{1})\le\Psi(X^{n},Y^{n})\le n\Psi(X_{1},Y_{1})$
%by monotonicity and subadditivity. Using the SFRL, we have $\Psi(X^{n},Y^{n})\le\log n+\log(I(X_1;Y_1)+1)+4$.
%However, it is uncertain whether $\Psi(X^{n},Y^{n})$ grows like a
%constant, $\log n$, or in between. ??????????????????????????????????????????????
%\end{rem}
\medskip

%--------------------------

%\section{Reverse SFRL}
%
%The following result is a direct consequence of SFRL
%\begin{prop}
%[Reverse strong functional representation lemma]\label{prop:rsfrl}
%Fix any joint distribution $\P_{XY}$ with $I(X;Y)<\infty$. There
%exists a conditional distribution $\P_{Z|X}$ such that 
%\[
%H(Y|X)\le I(X;Y)+\log(I(X;Y)+1)+4,
%\]
%and the lautum information
%\[
%L(Z;Y)\le\log(I(X;Y)+1)+4.
%\]
%Moreover, if $\left|\mathcal{X}\right|,\left|\mathcal{Y}\right|<\infty$,
%we can have $\left|\mathcal{Z}\right|\le\left|\mathcal{Y}\right|^{\left|\mathcal{X}\right|}$.
%\end{prop}
%Note that

%\section{Privacy Funnel}
%
%
\section{Acknowledgments}
The authors would like to thank the anonymous reviewers for their insightful remarks and for pointing out to us the Braverman-Garg paper. Their comments have helped improve the presentation of the results and their connections to previous work.
\appendix
%dummy comment inserted by tex2lyx to ensure that this paragraph is not empty%dummy comment inserted by tex2lyx to ensure that this paragraph is not empty

\subsection{Proof of Theorem \ref{thm:strongfrl}\label{subsec:strongfrl}}

Condition on the event $\{X=x\}$. First we show that $g_{X\to Y}(x,\,\{(\tilde{Y}_{i}, T_{i})\})$
 follows the distribution $\P_{Y|X}(\cdot|x)$. By the marking theorem
of the Poisson point process~\cite{kingman1992poisson}, $\{(\tilde{Y}_{i}, T_{i})\}$
is a Poisson point process over the product measure $\P_{Y}\times \mu$
(where $\mu$ is the Lebesgue measure on $[0,\infty)$). By the displacement
theorem~\cite{kingman1992poisson}, 
\[
\left\{ \left(\tilde{Y}_{i}\,,\,T_{i}\cdot\frac{d\,\P_{Y}}{d\,\P_{Y|X}(\cdot|x)}(\tilde{Y}_{i})\right)\right\} 
\]
is a Poisson point process over $\P_{Y|X}(\cdot|x)\times \mu$. Hence
\[
\min_{i}T_{i}\cdot\frac{d\,\P_{Y}}{d\,\P_{Y|X}(\cdot|x)}(\tilde{Y}_{i})\,\sim\,\Exp(1),
\]
and
\[
\tilde{Y}\left(\underset{i}{\arg\min}T_{i}\cdot\frac{d\,\P_{Y}}{d\,\P_{Y|X}(\cdot|x)}(\tilde{Y}_{i})\right)\,\sim\,\P_{Y|X}(\cdot|x),
\]
where we write $\tilde{Y}(k)=\tilde{Y}_k$. Now we bound $H(Y\,|\,\{(\tilde{Y}_{i}, T_{i})\})$. Let
\begin{align*}
\Theta &=\min_{i}T_{i}\cdot\frac{d\,\P_{Y}}{d\,\P_{Y|X}(\cdot|x)}(\tilde{Y}_{i}),\\
K &=\underset{i}{\arg\min}T_{i}\cdot\frac{d\,\P_{Y}}{d\,\P_{Y|X}(\cdot|x)}(\tilde{Y}_{i}),
\end{align*}
then $H(Y\,|\,\{(\tilde{Y}_{i}, T_{i})\})\le H(K)$. Conditioned on
$\Theta=\theta$, $\tilde{Y}_{K}\sim\P_{Y|X}(\cdot|x)$ and $\{(\tilde{Y}_{i}, T_{i})\}_{i\neq K}$
is a Poisson point process over the semidirect product measure
\[
\nu(A\times B)=\int_{A}\mu\left(B\,\cap\,\left[\theta\cdot\frac{d\,\P_{Y|X}(\cdot|x)}{d\,\P_{Y}}(y)\,,\,\infty\right)\right)d\P_{Y}(y).
\]
Note that $K-1=\left|\left\{ i:\,T_{i}<T_{K}\right\} \right|$. Hence $K-1$ conditioned on $\Theta=\theta$ and $\tilde{Y}_{K}=\tilde{y}$
follows the Poisson distribution with rate{\allowdisplaybreaks
\begin{align*} 
 \nu\left(\mathcal{Y} \times \left[0,\,T_{K}\right)\right)
 & =\nu\left(\mathcal{Y} \times \left[0,\,\theta\cdot\frac{d\,\P_{Y|X}(\cdot|x)}{d\,\P_{Y}}(\tilde{y})\right)\right)\\
 & =\int_{\mathcal{Y}}\mu\left(\left[0,\,\theta\cdot\frac{d\,\P_{Y|X}(\cdot|x)}{d\,\P_{Y}}(\tilde{y})\right)\,\cap\,\left[\theta\cdot\frac{d\,\P_{Y|X}(\cdot|x)}{d\,\P_{Y}}(y)\,,\,\infty\right)\right)d\P_{Y}(y)\\
 & =\theta\int_{\mathcal{Y}}\max\left\{ 0,\,\frac{d\,\P_{Y|X}(\cdot|x)}{d\,\P_{Y}}(\tilde{y})-\frac{d\,\P_{Y|X}(\cdot|x)}{d\,\P_{Y}}(y)\right\} d\P_{Y}(y)\\
 & \le\theta\int_{\mathcal{Y}}\frac{d\,\P_{Y|X}(\cdot|x)}{d\,\P_{Y}}(\tilde{y})\cdot d\P_{Y}(y)\\
 & =\theta\frac{d\,\P_{Y|X}(\cdot|x)}{d\,\P_{Y}}(\tilde{y}).
\end{align*}}
Therefore{\allowdisplaybreaks
\begin{align*}
\E\left[\log K\right] & =\E_{Y\sim\P_{Y|X}(\cdot|x)}\left[\int_{0}^{\infty}e^{-\theta}\E\left[\log K\,|\,\Theta=\theta,\,\tilde{Y}_{K}=Y\right]d\theta\right]\\
 & \le\E_{Y\sim\P_{Y|X}(\cdot|x)}\left[\int_{0}^{\infty}e^{-\theta}\log\left(\theta\frac{d\,\P_{Y|X}(\cdot|x)}{d\,\P_{Y}}(Y)+1\right)d\theta\right]\\
 & \le\E_{Y\sim\P_{Y|X}(\cdot|x)}\left[\log\left(\int_{0}^{\infty}e^{-\theta}\theta\frac{d\,\P_{Y|X}(\cdot|x)}{d\,\P_{Y}}(Y)d\theta+1\right)\right]\\
 & =\E_{Y\sim\P_{Y|X}(\cdot|x)}\left[\log\left(\frac{d\,\P_{Y|X}(\cdot|x)}{d\,\P_{Y}}(Y)+1\right)\right]\\
 & \le\E_{Y\sim\P_{Y|X}(\cdot|x)}\left[\max\left\{ \log\frac{d\,\P_{Y|X}(\cdot|x)}{d\,\P_{Y}}(Y)\,,0\right\} +1\right]\\
 & =D(\P_{Y|X}(\cdot|x)\,\Vert\,\P_{Y})-\E_{Y\sim\P_{Y|X}(\cdot|x)}\left[\min\left\{ \log\frac{d\,\P_{Y|X}(\cdot|x)}{d\,\P_{Y}}(Y)\,,0\right\} \right]+1\\
 & \le D(\P_{Y|X}(\cdot|x)\,\Vert\,\P_{Y})+e^{-1}\log e+1,
\end{align*}}
where the last line follows by the same arguments as in Appendix A in \cite{harsha2010communication}.
For $X\sim\P_{X}$,
\[
\E\left[\log K\right]\le I(X;Y)+e^{-1}\log e+1.
\]
By the maximum entropy distribution subject to a given $\E\left[\log K\right]$
(see Appendix~\ref{subsec:boundent}), we have
\begin{align*}
H(K) & \le I(X;Y)+e^{-1}\log e+2+\log\left(I(X;Y)+e^{-1}\log e+2\right)\\
 & \le I(X;Y)+\log\left(I(X;Y)+1\right)+e^{-1}\log e+2+\log\left(e^{-1}\log e+2\right)\\
 & <I(X;Y)+\log\left(I(X;Y)+1\right)+4.
\end{align*}

To prove the cardinality bound, first note that if $|\Xc|$, $|\Yc|$ are finite, then $\left|\mathcal{Z}\right|\le\left|\mathcal{Y}\right|^{\left|\mathcal{X}\right|}$ can be assumed to be finite since it is the number of
different functions $x\mapsto g_{X\to Y}(x,z)$ for different $z$.
To further reduce the cardinality, we apply Carath\'eodory's theorem on the $(|\Xc|(|\Yc|-1)+1)$-dimensional
vectors with entries $H(Y|Z=z)$ and $p(x,y|z)$ for $x\in \{1,\ldots,|\Xc|\}$,  $y\in \{1,\ldots,|\Yc|-1\}$; see~\cite{Ahlswede--Korner1975,Wyner--Ziv1976}.
The cardinality bound can be proved using Fenchel-Eggleston-Carath\'eodory
theorem~\cite{Eggleston1958,Rockafellar1970}.

\subsection{Proof of the Bound on Entropy in Theorem \ref{thm:strongfrl}\label{subsec:boundent}}
\textcolor{black}{The proof of the following proposition follows from the standard argument in maximum entropy distribution. It is included here for the sake of completeness.}
\begin{prop}
Let $\Theta\in\{1,2,\ldots\}$ be a random variable, then
\[
H(\Theta)\le\E\left[\log\Theta\right]+\log\left(\E\left[\log\Theta\right]+1\right)+1.
\]
\end{prop}
\begin{IEEEproof}
Let $q(\theta)=c\theta^{-\lambda}$ where $\lambda=1+1/\E\left[\log\Theta\right]$,
and $c>0$ such that $\sum_{\theta=1}^{\infty}q(\theta)=1$. Note
that
\[
\begin{aligned}\sum_{\theta=1}^{\infty}\theta^{-\lambda} & \le1+\int_{1}^{\infty}\theta^{-\lambda}d\theta=1+\frac{1}{\lambda-1}.\end{aligned}
\]
 Therefore
\[
\begin{aligned}H(\Theta) & \le\sum_{\theta=1}^{\infty}p_{\Theta}(\theta)\log\frac{1}{q(\theta)}\\
 & =\sum_{\theta=1}^{\infty}p_{\Theta}(\theta)\left(\lambda\log\theta-\log c\right)\\
 & =\lambda\E\left[\log\Theta\right]+\log\left(\sum_{\theta=1}^{\infty}\theta^{-\lambda}\right)\\
 & \le\lambda\E\left[\log\Theta\right]+\log\left(1+\frac{1}{\lambda-1}\right)\\
 & =\E\left[\log\Theta\right]+\log\left(\E\left[\log\Theta\right]+1\right)+1.
\end{aligned}
\]
\textcolor{black}{Operationally, we would use the optimal prefix-free code for the Zipf distribution $q(\theta)$ to encode $\Theta$.}

\end{IEEEproof}

\subsection{Proof of Proposition \ref{prop:lb_example}\label{subsec:lb_example}}

Let $k\in \{0,1,\ldots\}$, $V\in[0:\,2^{k}-1]$, 
\[
p_{V}(v)=\gamma^{-1}2^{k-\left\lceil \log(v+1)\right\rceil },
\]
where $\gamma=2^{k-1}(k+2)$, and let $X\sim\mathrm{Unif}[0:\,2^{k}-1]$ \textcolor{black}{independent of $V$}, and
$Y=(X+V)\;\text{mod}\;2^{k}$. Note that $\left|\left\{ v:\,\gamma p_{V}(v)>t\right\} \right|=\gamma p_{V}\left(\left\lfloor t\right\rfloor \right)$
for $t\ge0$. We have 
\begin{align*} 
& -\sum_{y\in\mathcal{Y}}\int_{0}^{1}\P_{X}\left\{ p_{Y|X}(y|X)\ge t\right\} \log\left(\P_{X}\left\{ p_{Y|X}(y|X)\ge t\right\} \right)dt\\
 &\qquad =-\sum_{y\in\mathcal{Y}}\int_{0}^{1}\left(2^{-k}\left|\left\{ v:\,p_{V}(v)\ge t\right\} \right|\right)\log\left(2^{-k}\left|\left\{ v:\,p_{V}(v)\ge t\right\} \right|\right)dt\\
 &\qquad =k-\int_{0}^{1}\left|\left\{ v:\,p_{V}(v)\ge t\right\} \right|\log\left|\left\{ v:\,p_{V}(v)\ge t\right\} \right|dt\\
 &\qquad =k-\int_{0}^{1}\gamma p_{V}\left(\left\lfloor \gamma t\right\rfloor \right)\log\left(\gamma p_{V}\left(\left\lfloor \gamma t\right\rfloor \right)\right)dt\\
 &\qquad =k-\sum_{v=0}^{2^{k}-1}p_{V}(v)\log\left(\gamma p_{V}\left(v\right)\right)dt\\
 &\qquad =k-\log\gamma+H(V).
\end{align*}
And
\[
I(X;Y)  =H(Y)-H(Y|X)=k-H(V).
\]
By Proposition \ref{prop:phi_lb},
\[
\Psi(X\to Y)  \ge k-\log\gamma+H(V)-\left(k-H(V)\right)=2H(V)-\log\gamma.
\]
One can check that
\[
H(V)  =\frac{1}{2}k+\log(k+2)-\frac{3}{2}+\frac{1}{k+2}.
\]
Hence
\[
I(X;Y)  =\frac{1}{2}k-\log(k+2)+\frac{3}{2}-\frac{1}{k+2} \le\frac{1}{2}k,
\]
and
\begin{align*}
\Psi(X\to Y) & \ge k+2\log(k+2)-3+\frac{2}{k+2}-\log\left(2^{k-1}(k+2)\right)\\
 & =\log(k+2)-2+\frac{2}{k+2}\\
 & \ge\log(I(X;Y)+1)-1.
\end{align*}

%
%\subsection{????????????????????}
%
%\[
%-\zeta\ln x+ax+b\ge e^{-\lambda x}
%\]
%
%\[
%a+b=e^{-\lambda}
%\]
%\[
%b=e^{-\lambda}-a
%\]
%
%\[
%-\frac{\zeta}{x}+a+\lambda e^{-\lambda x}
%\]
%\textbackslash{}zeta
%\[
%-\zeta+a+\lambda e^{-\lambda}=0
%\]
%\[
%\zeta=a+\lambda e^{-\lambda}
%\]
%
%\[
%\frac{\zeta}{x^{2}}+a-\lambda^{2}e^{-\lambda x}
%\]
%
%\[
%\frac{1}{x^{2}}\left(a+\lambda e^{-\lambda}\right)-\lambda^{2}e^{-\lambda x}\ge0
%\]
%
%\[
%\frac{1}{y^{2}}\zeta-e^{-y}\ge0
%\]
%\[
%\zeta\ge0.55
%\]
%
%\[
%-0.55\ln x+ax+b\ge e^{-\lambda x}
%\]
%
%\[
%-\zeta\E\left[\ln X\right]+a+b\ge\E\left[e^{-\lambda X}\right]
%\]
%
%\[
%0.55\E\left[-\ln X\right]+e^{-\lambda}\ge\E\left[e^{-\lambda X}\right]
%\]
%\[
%\begin{aligned} & \int_{1}^{\infty}\frac{1}{t}\left(\E\left[e^{-(t/n)X}\right]\right)^{n}dt\\
% & \le\int_{1}^{\infty}\frac{1}{t}\left(0.55\E\left[-\ln X\right]+e^{-t/n}\right)^{n}dt
%\end{aligned}
%\]

\bibliographystyle{IEEEtran}
\bibliography{ref,nit}

\end{document}

%% file: defns.tex
\usepackage{xspace}
\usepackage{bbm}
\input{mathlig}

\newcommand{\muspace}{\mspace{1mu}}

\DeclareRobustCommand{\scond}{\mathchoice{\muspace\vert\muspace}{\vert}{\vert}{\vert}}
\mathlig{|}{\scond}

\DeclareRobustCommand{\discint}{\mathchoice{\mspace{-1.5mu}:\mspace{-1.5mu}}{\mspace{-1.5mu}:\mspace{-1.5mu}}{:}{:}}
\mathlig{::}{\discint}

\newcommand{\Xc}{\mathcal{X}}
\newcommand{\Yc}{\mathcal{Y}}
\newcommand{\Zc}{\mathcal{Z}}

\newcommand{\Yt}{{\tilde{Y}}}

\def\e{\epsilon}

\DeclareMathOperator\E{\textsf{E}}
\let\P\relax
\DeclareMathOperator\P{\textsf{P}}
\newcommand{\Bern}{\mathrm{Bern}}

\newcommand{\Exp}{\mathrm{Exp}}

\newcommand{\U}{\mathrm{Unif}}

\def\textiid{i.i.d.\@\xspace}
\newcommand\iid{\ifmmode\text{ i.i.d. } \else \textiid \fi}

\def\mathllap{\mathpalette\mathllapinternal}
\def\mathllapinternal#1#2{%
  \llap{$\mathsurround=0pt#1{#2}$}}

\def\clap#1{\hbox to 0pt{\hss#1\hss}}
\def\mathclap{\mathpalette\mathclapinternal}
\def\mathclapinternal#1#2{%
  \clap{$\mathsurround=0pt#1{#2}$}}

\let\oldstackrel\stackrel
\renewcommand{\stackrel}[2]{\oldstackrel{\mathclap{#1}}{#2}}

\renewcommand{\hbar}{h\mathllap{\overline{\vphantom{h}\hphantom{\rule{4.6pt}{0pt}}}\mspace{0.77mu}}}

\catcode`~=11 % make LaTeX treat tilde (~) like a normal character
\newcommand{\urltilde}{\kern -.06em\lower -.06em\hbox{~}\kern .02em}
\catcode`~=13 % revert back to treating tilde (~) as an active character

%% file: mathlig.tex
\catcode`@=11
\chardef\mathlig@atcode\count255

\def\actively#1#2{\begingroup\uccode`\~=`#2\relax\uppercase{\endgroup#1~}}
\def\mathlig@gobble{\afterassignment\mathlig@next@cmd\let\mathlig@next= }
\def\mathlig@delim{\mathlig@delim}
\def\mathlig@defcs#1{\expandafter\def\csname#1\endcsname}
\def\mathlig@let@cs#1#2{\expandafter\let\expandafter#1\csname#2\endcsname}
\def\mathlig@appendcs#1#2{\expandafter\edef\csname#1\endcsname{\csname#1\endcsname#2}}

\def\mathlig#1#2{\mathlig@checklig#1\mathlig@end\mathlig@defcs{mathlig@back@#1}{#2}\ignorespaces}

\def\mathlig@checklig#1#2\mathlig@end{%
 \expandafter\ifx\csname mathlig@forw@#1\endcsname\relax
 \expandafter\mathchardef\csname mathlig@back@#1\endcsname=\mathcode`#1%
 \mathcode`#1"8000\actively\def#1{\csname mathlig@look@#1\endcsname}%
 \mathlig@dolig#1\mathlig@delim
\fi
\mathlig@checksuffix#1#2\mathlig@end
}

\def\mathlig@checksuffix#1#2\mathlig@end{%
\ifx\mathlig@delim#2\mathlig@delim\relax\else\mathlig@checksuffix@{#1}#2\mathlig@end\fi
}
\def\mathlig@checksuffix@#1#2#3\mathlig@end{%
\expandafter\ifx\csname mathlig@forw@#1#2\endcsname\relax\mathlig@dosuffix{#1}{#2}\fi
\mathlig@checksuffix{#1#2}#3\mathlig@end
}

\def\mathlig@dosuffix#1#2{%
\mathlig@appendcs{mathlig@toks@#1}{#2}%
\mathlig@dolig{#1}{#2}\mathlig@delim
}

\def\mathlig@dolig#1#2\mathlig@delim{%
 \mathlig@defcs{mathlig@look@#1#2}{%
 \mathlig@let@cs\mathlig@next{mathlig@forw@#1#2}\futurelet\mathlig@next@tok\mathlig@next}%
 \mathlig@defcs{mathlig@forw@#1#2}{%
  \mathlig@let@cs\mathlig@next{mathlig@back@#1#2}%
  \mathlig@let@cs\checker{mathlig@chck@#1#2}%
  \mathlig@let@cs\mathligtoks{mathlig@toks@#1#2}%
  \expandafter\ifx\expandafter\mathlig@delim\mathligtoks\mathlig@delim\relax\else
  \expandafter\checker\mathligtoks\mathlig@delim\fi
  \mathlig@next
 }%
 \mathlig@defcs{mathlig@toks@#1#2}{}%
 \mathlig@defcs{mathlig@chck@#1#2}##1##2\mathlig@delim{%
  \ifx\mathlig@next@tok##1%
   \mathlig@let@cs\mathlig@next@cmd{mathlig@look@#1#2##1}\let\mathlig@next\mathlig@gobble
  \fi 
  \ifx\mathlig@delim##2\mathlig@delim\relax\else
   \csname mathlig@chck@#1#2\endcsname##2\mathlig@delim
  \fi
 }%
 \ifx\mathlig@delim#2\mathlig@delim\else
  \mathlig@defcs{mathlig@back@#1#2}{\csname mathlig@back@#1\endcsname #2}%
 \fi
}%

\catcode`@\mathlig@atcode